\documentclass[preprint,12pt]{elsarticle}
\usepackage[utf8]{inputenc}
\usepackage[T1]{fontenc}
\usepackage{mathptmx}
\usepackage{amsmath,amssymb,amsfonts}
\usepackage{siunitx}
\usepackage{graphicx}
\graphicspath{{Figures/}}
\usepackage{booktabs}
\usepackage{multirow}
\usepackage{tabularx}
\usepackage{caption}
\usepackage{subcaption}
\usepackage{geometry}
\usepackage{etoolbox}
\usepackage{float}
\usepackage{xcolor}
\usepackage[section]{placeins}  % keep floats within sections
\usepackage[version=4]{mhchem}
\usepackage[document]{ragged2e}

\usepackage{hyperref}        % load near-last
\usepackage[capitalise]{cleveref}  % load AFTER hyperref
\journal{Journal of Nuclear Materials}
\biboptions{numbers,sort&compress}   % <-- key line
\colorlet{red}{black}

\begin{document}

\begin{frontmatter}

\title{Ductility Design Rules for Tungsten based Refractory  High Entropy Alloys from Sparse Experimental Datasets}

\author{
Avik Mahata\thanks{Corresponding author: \texttt{mahataa@merrimack.edu}} and Nick Beaver \\
}

\cortext[cor1]{Corresponding author}

\address {Department of Mechanical and Electrical Engineering, Merrimack College, North Andover, MA, 01845, USA}

\begin{abstract}

Tungsten-based refractory high-entropy alloys (RHEAs) are promising materials for fusion environments, but they often remain brittle at room temperature due to the intrinsically high ductile-to-brittle transition temperature (DBTT) of tungsten. To explore alloying strategies that may reduce room-temperature brittleness, we compiled a curated dataset of experimentally reported tungsten-containing alloys and constructed composition-based, physics-informed descriptors as inputs for machine learning. Three classifiers were evaluated using nested cross-validation, and a support vector classifier (SVC) demonstrated the most reliable generalization performance for this sparse dataset. Model interpretation using Shapley additive explanations identified the exchange--correlation parameter, valence electron concentration, pressure field, and electronegativity mismatch as the most influential features governing the ductile--brittle classification boundary. Synthetic compositions generated within the convex interpolation domain of the training data were evaluated and visualized on pseudo-ternary diagrams to visualize predicted ductility trends across composition space. The model predicts that moderate additions of Ti, Ni, and Co increase the likelihood of room-temperature ductility, whereas high Cr contents shift alloys toward brittle behavior. These trends are consistent with experimentally reported observations in tungsten-based systems, including studies demonstrating Ti-assisted ductility improvements through solid-solution and electronic effects and reports of embrittlement in Cr-rich refractory alloys. \textcolor{red}{Broad agreement between the DFT-derived elastic descriptors and the SVC decision-function margins further supports the physical relevance of the inferred classification boundary.} While the available mechanical dataset remains limited and heterogeneous, the quantitative agreement between learned trends and published experimental behavior supports the physical validity of the inferred decision boundary. This work establishes an interpretable, physics-informed screening framework tailored to sparse tungsten-alloy datasets, enabling identification of candidate alloys for targeted simulation and experimental validation in fusion-relevant environments.

\end{abstract}


\end{frontmatter}

\raggedright

\section{Introduction}

Tungsten has long been used in extreme thermomechanical environments because of its very high melting point of 3422~$^\circ$C, strong creep resistance, and excellent high-temperature stability \cite{lassner1999,rieth2013, webb2019overview, das2014thermo}. Its high thermal conductivity \cite{lee2012thermal}, low sputtering yield \cite{eckstein1991sputtering}, chemical inertness to hydrogen isotopes \cite{venhaus2001behavior}, and minimal tritium retention \cite{haasz1997deuterium} make it one of the most promising plasma-facing material candidates for future fusion reactors \cite{philipps2011,federici2003}. However, ITER-grade polycrystalline tungsten exhibits an intrinsically high DBTT, typically near 400~$^\circ$C, along with a recrystallization temperature of approximately 1200~$^\circ$C \cite{wronski1965ductile, taezl2018, lassila1991ductile}. This high DBTT severely limits toughness during reactor shutdown and handling, while neutron irradiation further increases brittleness by generating lattice defects, reducing thermal conductivity, and promoting crack initiation under cyclic and steady heat loads \cite{zinkle2012,riethImpact2018}. Consequently, lowering the DBTT while maintaining high recrystallization resistance and high-temperature strength has become a central objective. Although conventional grain-boundary, deformation, and particle-based dispersion strengthening have been explored, excessive second-phase additions often degrade thermal conductivity or induce grain-boundary embrittlement \cite{riethImpact2018,liu2019}.

In fusion environments, plasma-facing tungsten components operate within constrained service temperature windows that are further narrowed by neutron irradiation. Irradiation-induced defect accumulation is known to significantly increase the ductile-to-brittle transition temperature in tungsten-based materials, thereby shifting the minimum allowable operating temperature upward \cite{Hu2016IrrHardening, Abernethy2019DBTT, Terentyev2021DBTT}. In addition to embrittlement, neutron exposure degrades thermal conductivity and promotes transmutation-driven compositional changes, which further complicate alloy design \cite{Cui2018ThermalCond, Lang2023Transmutation}. These coupled effects highlight that ductility optimization alone is insufficient for fusion qualification. 

To address these limitations, W-based refractory high-entropy alloys (RHEAs) have emerged as promising candidates. By incorporating multiple principal elements into a single BCC matrix, RHEAs can modify electronic structure, lattice resistance, and elastic interactions in ways that may reduce the ductile-to-brittle transition temperature while retaining high-temperature strength. Additionally, the chemical complexity of these alloys has been shown in some systems to enhance phase stability and irradiation tolerance. 

However, it should be emphasized that W-based high-entropy alloys do not inherently solve all fusion-relevant materials challenges. In particular, compositionally complex W-HEAs generally exhibit lower intrinsic thermal conductivity than pure tungsten due to enhanced electron and phonon scattering arising from chemical disorder. For example, Shi et al. reported that the electrical resistivity of a WTaCrVTi high-entropy alloy is approximately an order of magnitude higher than that of pure W, corresponding to a significantly reduced thermal conductivity under Wiedemann–Franz analysis \cite{Shi2024JNM}. Furthermore, irradiation substantially degrades the thermal transport of pure tungsten itself. Cui et al. demonstrated that the thermal conductivity of ion-irradiated W decreases to roughly one-third of its pristine value above 0.2 dpa \cite{Cui2018JNM}, and Sina et al. reported up to a 50\% reduction in thermal diffusivity in highly irradiated tungsten exposed to spallation conditions \cite{Sina2024JNM}. 

These observations indicate that the design problem is inherently multi-objective rather than single-property optimization. While plasma-facing components may experience transient surface temperatures exceeding 2000~$^\circ$C, the present study does not propose W-RHEAs as universal replacements for pure W in all high-temperature regimes. Instead, the objective is to explore compositional strategies that mitigate room-temperature brittleness and reduce DBTT within constrained operating windows. The present work therefore focuses on room-temperature ductility as an initial screening criterion, while recognizing that melting temperature, thermal transport retention, irradiation-induced property evolution, and activation constraints must ultimately be considered together in future multi-objective alloy optimization frameworks.

Historically, efforts to reduce the brittleness of tungsten progressed from empirical alloy development to physics based understanding. Early metallurgical work showed that adding rhenium (Re) improves ductility and lowers the DBTT. This phenomenon became known as the rhenium effect \cite{stephens1980, romaner2010effect, gornostyrev1991nature}. Microstructural engineering also played an important role. Potassium doping in drawn tungsten wire increases toughness by creating stable pore structures that restrain grain boundary motion \cite{riesch2016development, nogami2019improvement, sheng2012mechanical}. Grain refinement methods likewise demonstrated that careful control of grain boundary morphology is critical for improving toughness \cite{dekloe2008}. Recent work on refractory multi principal element alloys (MPEAs) has shown that specific alloying additions can change electronic structure and bonding. These changes shift the balance between brittle and ductile behavior \cite{riethImpact2018, liu2019, tandoc2023mining, tandoc2023mining}. First principles investigations of Al and Cu substitution in MoNbTaW have confirmed this idea. These studies found that modifying charge localization, the width of the d band, and the valence electron concentration can promote ductility while maintaining thermodynamic stability \cite{woodcox2025enhancing, chen2024structure, qi2024integrated}.

Although such advances have provided critical mechanistic insight, a complete understanding of tungsten brittleness remained elusive until the advent of high-fidelity atomistic modeling. Density Functional Theory (DFT) established that the intrinsic origin of brittleness in BCC tungsten arises from the high lattice resistance to the motion of $1/2\langle111\rangle$ screw dislocations, which must nucleate kink pairs to glide \cite{vitek1968}. DFT has also quantified grain-boundary embrittlement due to impurity segregation (O, P, S) and revealed how solutes such as Re alter bonding to improve cohesion \cite{cannon2005}. Molecular Dynamics (MD) \cite{armstrong2014, petersson2023molecular, alivaliollahi2023effect} and Monte Carlo (MC) simulations \cite{cereceda2015linking} have extended this static picture by capturing the competition between dislocation emission and crack propagation at grain boundaries, as well as solute segregation behavior under irradiation. However, despite the accuracy of these atomistic tools, they remain computationally prohibitive for exploring the vast compositional design space of W-based refractory HEAs. This is where machine learning (ML) and artificial intelligence (AI) provide a transformative opportunity.  ML has already demonstrated broad impact across scientific fields—from drug discovery \cite{jumper2021} to high-throughput compositional screening \cite{jain2013}—and offers a pathway to extract design rules from sparse and heterogeneous materials datasets. Yet, for tungsten-containing HEAs, the available experimental mechanical property data remain limited, inconsistently reported, and insufficient on their own to guide alloy design without a unifying, physics-informed modeling framework.

In this work, we address this challenge by compiling a curated dataset of tungsten-containing refractory HEAs drawn from several experimental sources. We focus on alloys for which room-temperature ductility or brittleness has been clearly reported. To approximate aspects of bonding, deformation, and stability, we compute a set of composition-based descriptors. These include valence electron concentration, atomic size mismatch, enthalpy of mixing, averaged elastic moduli, and average exchange–correlation parameters. Each descriptor is chosen for interpretability and physical relevance. These rule of mixtures features allow for some interpolation across composition space while avoiding extrapolation outside the domain of known alloy behavior. To identify which learning framework best separates ductile from brittle W-HEAs, we evaluate several machine learning classifiers, including random forests, logistic regression, and support vector classifiers. A nested cross validation procedure is used to limit overfitting in the small dataset. The support vector classifier provides the most reliable generalization and is therefore selected for final inference. Shapley value analysis gives an interpretable description of the decision boundary and highlights the importance of electronic descriptors such as the average BCC exchange–correlation parameter, pressure field, valence electron concentration, and electronegativity mismatch. To visualize how alloying affects ductility within the high dimensional design space, we generate synthetic compositions that remain within the extrema of the training data and evaluate them using the trained SVC. The decision function is projected onto pseudo ternary diagrams, which reveal clear trends. Titanium and nickel additions increase the likelihood of ductility, while chromium and hafnium tend to decrease it. One dimensional compositional analyses support these observations. This approach creates an interpretable and data driven framework for identifying alloying strategies that lower the DBTT in tungsten-containing refractory HEAs. By combining curated experiments with physics based descriptors, robust model selection, and clear visualization of learned design rules, this work supports the discovery of ductile and fusion-relevant W-based alloys.

\section{Computational Methodology}

\subsection{Dataset Preparation}

The dataset used in this study was curated from experimental reports on tungsten-based RHEAs containing W and related refractory elements. The curated dataset was combined with the much larger datasets [\cite{li2023database}, \cite{borgexpanded},\cite{detor2022refractory}] and filtered to only include alloys that contained W to study the ductility of the alloy class of interest. The data was restricted to tests performed at room temperature ($\leq 30^\circ$C) only, since the majority of reference data in the curated datasets reported mechanical properties for sparse temperatures (i.e. $30^\circ$C, $600^\circ$C, $1000^\circ$C) to examine the alloy response across a wide range of temperatures. If an alloy was ductile at room temperature, it was considered to be a useful alloy in the sense that the true goal behind lowering the DBTT of W alloys is to mitigate the risk of component failure from the lack of toughness of brittle alloys during reactor down time \cite{reiser2017}. For much the same reason, our prediction task was chosen to be a binary classification task instead of a regression task, since we only sought to find the decision boundary between usable and non-usable compositions with respect to ductility. 

DBTT is not an intrinsic property of a material, but is dependent on test criteria \cite{butler2018mechanisms} that represents a temperature-dependent transition between types of mechanical response rather than a single-point property. However, the scientific literature establishes that RT ductility and DBTT share a common mechanistic origin. Ductility in these alloys rely on thermally-activated screw dislocation motion via kink-pair nucleation \cite{lu2021relative, gumbsch1998controlling}. Critically, the tungsten literature demonstrates that metallurgical interventions improving RT ductility, including Re alloying, ZrC dispersion, thermomechanical processing, also reduce DBTT \cite{ren2018methods, xie2015extraordinary, li2012dislocation}. Recent work on refractory HEAs gives more evidence for this correlation. RT-ductile TiZrHfNbTa exhibits a DBTT of 247 K while RT-brittle VNbMoTaW has a DBTT of 627 K, with the difference arising from the same dislocation core structures controlling  both properties \cite{tsuru2024intrinsic}. Given the scarcity of comprehensive DBTT data for the W-RHEA design space, RT ductility classification represents a physically-justified early-stage screening metric that probes the same fundamental mechanisms controlling DBTT.

Each entry in the final dataset includes experimentally measured mechanical properties such as yield strength and total elongation at room temperature. However, the mechanical test performed on each alloy in the dataset was not the same in every case. The lack of sufficient, standardized fracture and DBTT-related data has been identified as a key barrier to RHEA alloy design for fusion applications~\cite{hatler2025path}. Many papers report only the compressive properties of a fabricated alloy, but the percentage of strain at failure is higher in compressive tests than in tensile tests for the same alloy due to the tensile-compressive asymmetry in W alloys~\cite{butler2018mechanisms}. To allow for both tensile and compressive strains to be used in one training set, we defined a cutoff for what would be considered a ductile alloy for both cases. If an alloy had a tensile elongation at failure of $\geq 5\%$, it was taken to be a ductile alloy at room temperature, which is a threshold that has been used to study DBTT behavior in W-Re~\cite{raffo1969yielding}. To account for the difference in response between tensile and compressive regimes, we used a compressive strain at failure of $\geq 10\%$ to define ductile alloys in the cases where only compressive data was reported. 

\textcolor{red}{Because no standardized room-temperature ductility criterion exists across the W-RHEA literature, a unified classification framework was required to combine tensile and compressive mechanical data reported by different studies. The tensile threshold of 5\% elongation is consistent with values commonly used when assessing ductility and DBTT-related behavior in tungsten-based alloys, while the corresponding compressive threshold of 10\% reflects the well-established tensile-compressive asymmetry observed in refractory BCC systems. Similar strain-based and ductility-screening criteria are commonly employed in refractory alloy design studies, particularly when direct fracture-toughness or DBTT measurements are unavailable~\cite{ouyang2023design,senkov2021generalization}. To evaluate the robustness of the adopted criterion, a sensitivity analysis was performed in Section~\ref{sec:validation}, where the complete nested cross-validation workflow was repeated using alternative compressive strain thresholds. The resulting performance metrics varied only modestly across the tested range, indicating that the principal conclusions of this work are not strongly dependent on the specific threshold selected. While a future dataset consisting entirely of standardized tensile, fracture-toughness, or direct DBTT measurements would provide a more rigorous basis for classification, the present approach enables physically motivated screening using the heterogeneous experimental data currently available for tungsten-containing refractory alloys.}

To improve transparency and traceability of the mechanical property data used for model training, a concise summary of the curated dataset is provided in \textbf{Table~1}. The complete dataset, including individual alloy compositions, reported mechanical properties, and original literature sources, is available in the associated GitHub repository (See Data Availability). This summary clarifies the classification criteria and scope of the dataset used for model development. The following section describes the computation of compositional descriptors derived from these alloys. The complete dataset, as well as all of the source data and the aggregation code, can be found in the Data Availability section of the related GitHub repository.

\begin{table}[H]
\centering
\caption{Summary of experimentally reported room-temperature mechanical data used for binary ductility classification.}
\begin{tabular}{lc}
\hline
Category & Value \\
\hline
Total number of alloys & 87 \\
Ductile alloys & 40 \\
Brittle alloys & 47 \\
Mechanical test types & Tension and Compression \\
Tensile ductility criterion & $\varepsilon_f \ge 5\%$ \\
Compressive ductility criterion & $\varepsilon_f \ge 10\%$ \\
Temperature range & $\le 30^\circ$C \\
\hline
\end{tabular}
\end{table}

\begin{figure}[H]
    \centering
    \includegraphics[width=1\textwidth]{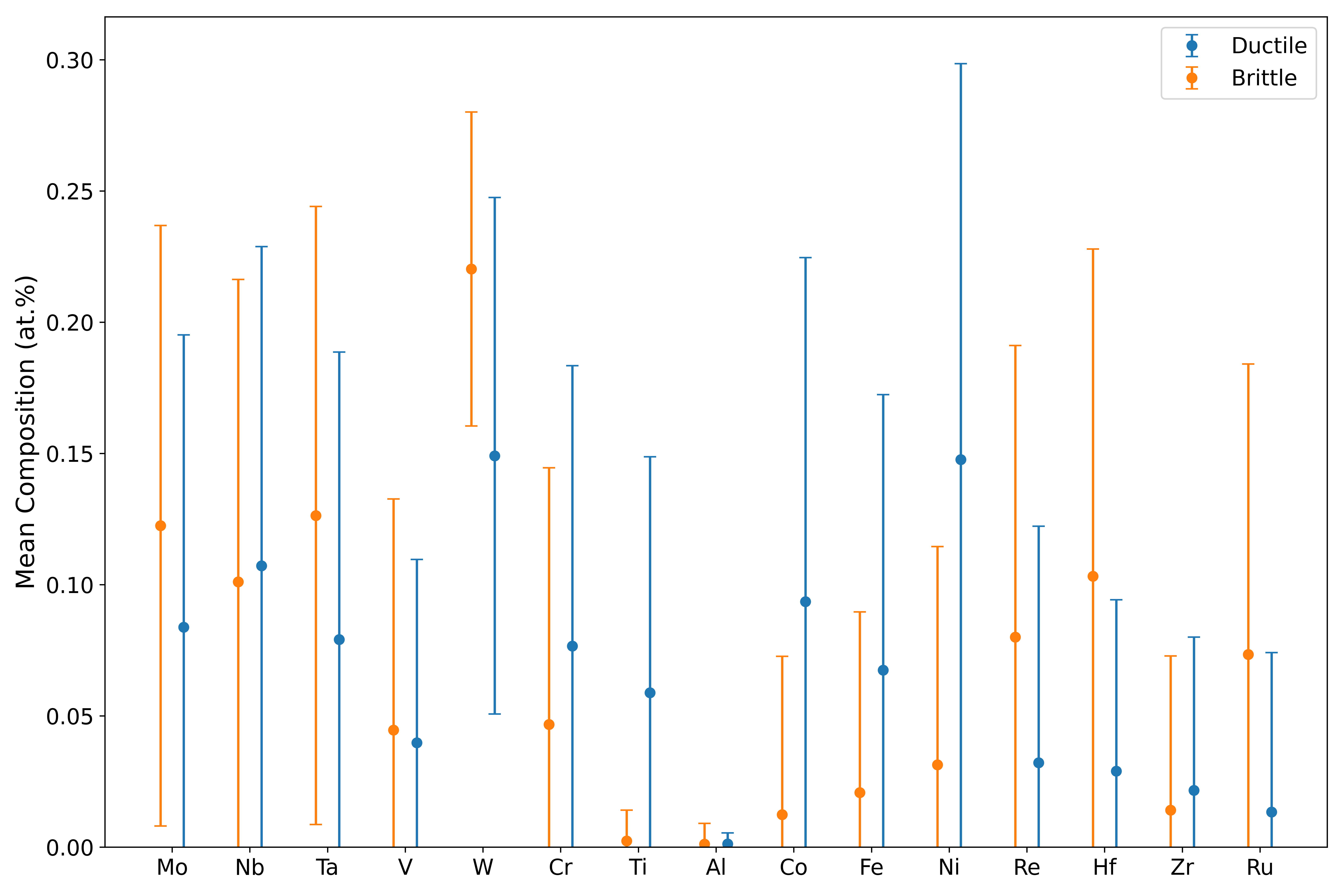}
    \caption{Average composition of the alloys in the dataset separated into room-temperature ductility classification. For each element in the dataset, the average composition for ductile (blue) and brittle (orange) alloys is shown alongside an error bar that is the standard deviation of composition}
    \label{fig:Bar}
\end{figure}

\textbf{Fig.~\ref{fig:Bar}} (referred to as \textbf{Fig.~1} in the main text) shows the average composition of each element in the dataset grouped by ductility class. As seen in \textbf{Fig.~1}, ductile alloys generally contain higher fractions of non-refractory elements and reduced amounts of W and Ta, indicating that incorporating more ductile-promoting solutes correlates with improved room-temperature toughness.

Even before any training was performed on the set, a clear trend in ductility is seen. Ductile alloys in our experimental dataset tended to have non-refractory elements alloyed in higher concentrations, and also tended to have less W and Ta. Even though it is possible to visualize the alloying strategies already present in the dataset, predicting unseen alloys and discovering alloy strategies requires a feature set with general, interpretable information about the alloys themselves. A popular approach in the field of HEAs has been to use rule-of-mixtures (ROM) calculations and other functions of composition to calculate average features of alloys. 

The most common ROM features are arithmetic means of a property (P) in terms of atomic fractions within a composition (c) $P = \sum_{i=1}^{n} c_i P_i$. Composition-based/averaged parameters have been shown to work for predicting the properties of HEAs, and are especially useful in the prediction of phase formation, high-throughput filtering of a candidate space \cite{ouyang2023design}, or as inputs for machine learning of the mechanical properties of high entropy alloys \cite{qi2024integrated}. ROM calculations act like crude estimations of physical properties, for example how Valence Electron Concentration (VEC) 
is empirically related to ductility and phase structure \cite{qi2024integrated}. Using approximate estimates of physics-informed features as inputs for machine learning has been shown to increase model performance on small materials datasets \cite{zhang2018strategy}. Some of the physics of the output (ductility in our case) may be compressed in average electronic or physical properties, which is crucial for small materials dataset that tend to underfit the training data \cite{zhang2018strategy}. Another benefit of using ROM parameters as inputs is that it gave us the ability to predict and visualize the ductility boundary of a wider range of alloys than just the small set we had, since we could interpolate in our feature space.

From each alloy composition, a set of interpretable compositional descriptors was calculated to serve as model input features. These parameters, summarized in \textbf{Table~\ref{tab:parameters}}, 

include averaged or variance-based features derived from the elemental properties. The formulation of each descriptor follows established thermodynamic and empirical mixing rules and represents scalar quantities that might approximate the overall structural, chemical, and mechanical tendencies of the alloy system. They were chosen to be cheap to calculate and only depend on the composition of the alloy and the properties of its constituents. Most of the parameters as well as the elemental data for ROM calculations were adapted into python code from \cite{martin2022heaps} and \cite{qi2024integrated}, and are included in the final repository.

\begin{table}[H]
\centering
\small
\renewcommand{\arraystretch}{1.4}
\caption{Parameters and calculation formulas used as compositional input features for the machine learning model. Each descriptor is computed from elemental fractions $c_i$ and corresponding elemental properties.}
\vspace{6pt}
\begin{tabular}{|c|l|l|}
\hline
\textbf{No.} & \textbf{Parameter} & \textbf{Calculation Formula} \\
\hline
1 & Atomic size difference \cite{martin2022heaps} &
$\displaystyle \delta = \sqrt{\sum_{i=1}^{n} c_i \left(1 - \frac{r_i}{\bar{r}}\right)^2}$ \\
\hline
2 & Average melting point \cite{martin2022heaps} &
$\displaystyle T_m = \sum_{i=1}^{n} c_i T_{m_i}$ \\
\hline
3 & Enthalpy of mixing \cite{martin2022heaps} &
$\displaystyle \Delta H^{m} =
\sum_{i=1}^{n} \sum_{\substack{j=1 \\ j \ne i}}^{n}
4\, \Delta H_{ij}^{m}\, c_i c_j$ \\
\hline
4 & Ideal mixing entropy \cite{martin2022heaps} &
$\displaystyle \Delta S_{\mathrm{mix}} = -k_B \sum_{i=1}^{n} c_i \ln(c_i)$ \\
\hline
5 & Pairwise electronegativity mismatch \cite{qi2024integrated} &
$\displaystyle \delta_{\chi} =
\frac{\sum_{i=1}^{n} \sum_{\substack{j=1 \\ j \neq i}}^{n}
c_i c_j \left| \frac{\chi_i - \chi_j}{\bar{\chi}} \right|}
{\sum_{i=1}^{n} \sum_{\substack{j=1 \\ j \neq i}}^{n} c_i c_j}$ \\
\hline
6 & Valence electron concentration (VEC) \cite{martin2022heaps} &
$\displaystyle \mathrm{VEC} = \sum_{i=1}^{n} c_i\, \mathrm{Valency}_i$ \\
\hline
7 & Average Young's modulus \cite{qi2024integrated} &
$\displaystyle E = \sum_{i=1}^{n} c_i E_i$ \\
\hline
8 & Average shear modulus \cite{qi2024integrated} &
$\displaystyle G = \sum_{i=1}^{n} c_i G_i$ \\
\hline
9 & Average bulk modulus \cite{qi2024integrated} &
$\displaystyle B = \sum_{i=1}^{n} c_i B_i$ \\
\hline
10 & Average Poisson's ratio \cite{qi2024integrated} &
$\displaystyle \nu = \sum_{i=1}^{n} c_i \nu_i$ \\
\hline
11 & Average exchange–correlation parameter \cite{johnson2023universal} &
$\displaystyle r_s = \sum_{i=1}^{n} c_i\, r_{s_i}$ \\
\hline
12 & Pressure field \cite{qi2024integrated} &
$\displaystyle P = \frac{G (1 + \nu)}{3 \pi (1 - \nu)}$ \\
\hline
13 & Geometric strain \cite{qi2024integrated} &
$\displaystyle \frac{E_{2}}{E_{0}} \equiv
\sum_{i} \sum_{j>i}
\frac{c_{i} c_{j} \left| r_{i} + r_{j} - 2\bar{r} \right|^{2}}{(2\bar{r})^{2}}$ \\
\hline
14 & Pugh ratio \cite{pugh1954xcii} &
$\displaystyle \mathrm{PR} = \frac{G}{B} = \frac{G_{\mathrm{AVG}}}{B_{\mathrm{AVG}}}$ \\
\hline
15 & Residual strain \cite{qi2024integrated} &
$\displaystyle \sqrt{\langle \varepsilon^{2} \rangle} \equiv
\sum_{i=1}^{N} c_i \varepsilon_i^{2}, \quad
\varepsilon_i =
\frac{\sum_{j=1}^{N} \omega_{ij} c_j}{\sum_{k=1}^{N} A_{ik} c_k}
- \frac{4\pi \eta_{\mathrm{ideal}}}{N_i \sum_{k=1}^{N} A_{ik} c_k}$ \\
\hline
\end{tabular}

\vspace{6pt}
\begin{flushleft}
\footnotesize
\textbf{Key:}
$c_i$ -- atomic fraction of element $i$;\;
$n$ -- number of constituent elements;\;
$r_i$ -- atomic radius, $\bar{r}=\sum_i c_i r_i$ mean atomic radius;\;
$T_{m_i}$ -- elemental melting point;\;
$\Delta H_{ij}^{m}$ -- binary mixing enthalpy of $i$--$j$ (Miedema);\;
$k_B$ -- Boltzmann constant;\;
$\chi_i$ -- Pauling electronegativity, $\bar{\chi}$ its mean;\;
$\mathrm{Valency}_i$ -- number of valence electrons;\;
$E_i,\,G_i,\,B_i,\,\nu_i$ -- elemental Young's, shear, bulk moduli and Poisson's ratio ($G_{\mathrm{AVG}},B_{\mathrm{AVG}}$ their composition-weighted averages);\;
$r_{s_i}$ -- elemental exchange--correlation (Wigner--Seitz density) parameter, with $\tfrac{4}{3}\pi r_s^3 = 1/\rho_0$ and $\rho_0$ the interstitial electron density \cite{johnson2023universal};\;
$P$ -- pressure field;\;
$E_2/E_0$ -- geometric strain;\;
$\varepsilon_i$ -- per-element residual strain, where $\omega_{ij},\,A_{ik}$ are pairwise geometric weighting terms, $\eta_{\mathrm{ideal}}$ the ideal packing fraction, and $N_i$ the coordination number (after \cite{qi2024integrated}).
\label{tab:parameters}
\end{flushleft}
\end{table}

\subsection{Machine Learning Framework}

\begin{figure}[H]
    \centering
    \includegraphics[width=1.0\textwidth]{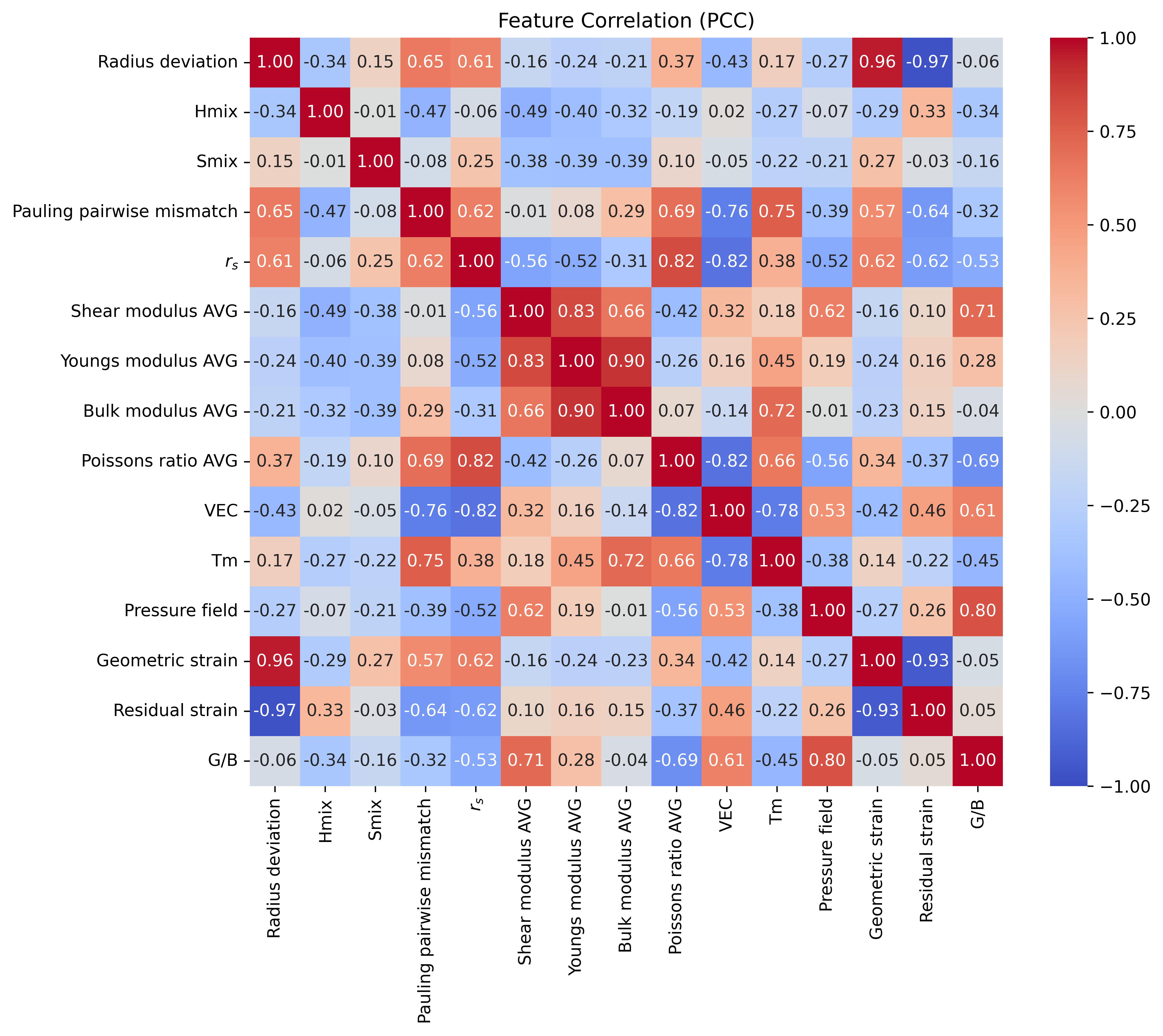}
    \caption{Pearson correlation coefficient (PCC) matrix of the computed compositional descriptors. The correlation coefficient between each pair of features is shown with a heatmap where red represents positive correlations, blue represents negative correlations, and soft/white colors represent a lack of correlation.Highly correlated features were pruned prior to model training.} 
    \label{fig:PCC}
\end{figure}

After feature computation, redundant or highly correlated parameters were identified through Pearson correlation coefficient (PCC) analysis to mitigate the effects of collinearity before training. The heatmap of coefficients is shown in \textbf{Fig.~\ref{fig:PCC}}. 

The Young's modulus, geometric strain, and residual strain features were removed since they were correlated to other features with a coefficient of $\geq 0.9$. Another benefit of removing some of the features is to reduce the dimensionality of the input space and thereby lower the degrees of freedom of the models used, which is important for increasing generalization (reducing overfitting) of models in physical datasets \cite{zhang2018strategy}. The refined feature set served as input for 3 classifiers commonly used on small datasets: a random forest (RF), logistic regression (LR), and SVC. Random forests are ensembles of decision trees, each trained on a bootstrap sample of the data. At each node split, only a random subset of features is considered, and the split that maximizes information gain among those candidates is chosen. They generalize well because averaging the decisions of many smaller trees trained on subsets of the data trees reduces variance while maintaining low bias. Logistic regression models the log-odds of class membership as a linear function of the input features, applying a sigmoid transformation to produce class probabilities. The decision boundary is a linear hyperplane in feature space, found by maximizing the sum of log-likelihoods over all training samples. Support vector machines find the boundary (which can be linear or nonlinear) that maximizes the distance between the boundary and each class. If the classes are not fully separable in the feature space, regularization parameters are used that balance the margin requirements with classification accuracy. When training any model, it is important to take into account the tradeoff between how well the training data is fit (bias) and how well the model generalizes to test data (variance). Well-optimized models tend to be in a sweet spot of complexity where they do not overfit (too many parameters, as in a very high-degree polynomial on linear data) nor underfit (not enough parameters, as in  a linear model on quadratic data). Each of the three models had a series of hyperparameters that needed to be fit to find the most optimal balance between bias and variance. The hyperparameter search space along with the chosen values for all three models are stated in \textbf{Table}~\ref{tab:hyperparameters}.

{\color{red}
\begin{table}[H]
\centering
\caption{Hyperparameter search spaces and selected values for the three classifiers, tuned via the inner cross-validation loop.}
\label{tab:hyperparameters}
\begin{tabular}{llll}
\hline
\textbf{Algorithm} & \textbf{Hyperparameter} & \textbf{Search space} & \textbf{Selected} \\
\hline
\multirow{4}{*}{Random Forest}
 & n\_estimators      & \{50, 100, 500\}            & 50 \\
 & max\_depth         & \{2, 3, 5, None\}          & 2 \\
 & min\_samples\_split & \{2, 3, 5\}               & 2 \\
 & min\_samples\_leaf  & \{1, 3, 5\}               & 1 \\
\hline
\multirow{4}{*}{Logistic Regression}
 & C        & \{0.001, 0.01, 0.1, 1, 10, 100\} & 10 \\
 & penalty  & \{l1, l2\}                        & l1 \\
 & solver   & \{liblinear\}                     & liblinear \\
 & max\_iter & \{1000\}                         & 1000 \\
\hline
\multirow{3}{*}{Support Vector Classifier}
 & C      & \{0.01, 0.1, 1, 10, 100\}              & 100\\
 & kernel & \{linear, rbf\}                         & rbf\\
 & gamma  & \{scale, auto, 0.001, 0.01, 0.1, 1\}    & 0.01\\
\hline
\end{tabular}
\end{table}

}

\begin{figure}[H]
    \centering
    \includegraphics[width=1\textwidth]{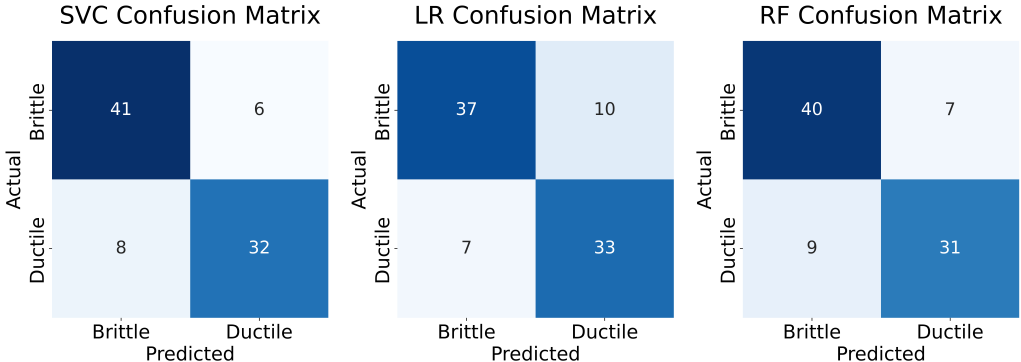}
    \caption{Aggregated confusion matrices from outer folds of validation. In each matrix, clockwise from the top left square, the true negatives, false positives, false negatives, and true positives are shown. Each model's counts for these metrics are counted across all validation folds.}
    \label{fig:confusion_matrices}
\end{figure}

\textcolor{red}{Nested, stratified k-fold cross-validation (5 inner folds, 10 outer folds) was performed to tune each model and find the best of the three for inference.} This strategy was shown to be very important to get an unbiased estimate of model generalization on small datasets \cite{cawley2010over}.  Each model was evaluated on how well it could distinguish between ductile and brittle alloys at room temperature. This methodology enabled interpretable, data-driven discovery of design rules for tungsten-based RHEAs by directly relating physically meaningful parameters to mechanical performance.\cite{talignani2022review}. The summary of the model performance is shown below in \textbf{Table~\ref{tab:model_performance}}. Confusion matrices (\textbf{Fig.~\ref{fig:confusion_matrices}}) were constructed by aggregating predictions from all outer folds of the nested cross-validation. Since each sample appears in exactly one outer test fold, these matrices provide an unbiased estimate of model performance on the complete dataset, with every sample predicted under conditions where it was held out from training. There was not a significant class imbalance of alloys (47 ductile, 40 brittle) according to our criteria for classification. 
However, this is still a small dataset, which creates challenges in learning and interpretation such as a possible sensitivity to individual data points and relative imbalance in element representation. The use of ROM features as inputs should reduce this bias somewhat, but it is important to note that the results of the study should only be interpreted as interpolative trends within the sampled compositional space rather than definitive design rules.

\subsubsection{Model Validation}
\label{sec:validation}

The most robust metric to use to evaluate model performance was F1 score, since it is defined as the harmonic mean of precision and recall and because we did not choose to make a distinction between the cost of false positives and false negatives. The results from the model selection process led us to use support vector classification with an rbf kernel for use on inference of the entire dataset, since this model showed the most balanced performance on unseen data during cross validation. This matched expectations, since it is a simple model that typically performs well on small datasets and high dimensional input space. 

\begin{table}[H]
\centering
\caption{Cross-validation performance metrics for classification models.}
\vspace{6pt}
\begin{tabular}{lcccc}
\hline
Model & Balanced Accuracy & Precision & Recall & F1 \\
\hline
Random Forest & 0.8150 ± 0.0989 & 0.8705 ± 0.1668 & 0.7750 ± 0.2077 & 0.7892 ± 0.1223 \\
Logistic Regression & 0.8075 ± 0.1360 & 0.7800 ± 0.1579 & 0.8250 ± 0.1953 & 0.7911 ± 0.1544 \\
Support Vector Machine& 0.8375 ± 0.1315 & 0.8650 ± 0.1817 & 0.8000 ± 0.1500 & 0.8234 ± 0.1434 \\
\hline
\end{tabular}

\label{tab:model_performance}
\end{table}

\textcolor{red}{Sensitivity analysis (\textbf{Table} \ref{tab:threshold_sensitivity}) the threshold was performed to verify our approach. Holding the tensile criterion fixed at 5\% and varying the compressive threshold over 5, 8, 10, 12, and 15\%, we repeated the full pipeline including per-threshold hyperparameter re-tuning via nested cross-validation for the SVC model. Balanced accuracy (the fairest cross-threshold comparison since precision, recall, and F1 shift mechanically with the changing class balance) remained within 0.79–0.83 with no systematic dependence on the cutoff, indicating the model's discriminative performance does not hinge on the 10\% choice.}

\begin{table}[H]
\centering
\caption{Sensitivity of SVC performance to the ductility strain threshold. The full nested cross-validation pipeline, including per-threshold hyperparameter re-tuning, was repeated for each threshold ($n=87$ alloys throughout). The 10\% criterion adopted in this work is shown in bold.}
\label{tab:threshold_sensitivity}
\begin{tabular}{cccccc}
\hline
\textbf{Threshold (\%)} & \textbf{Ductile ($n$)} & \textbf{Balanced acc.} & \textbf{F1} & \textbf{Precision} & \textbf{Recall} \\
\hline
5  & 49 & 0.791 & 0.838 & 0.786 & 0.898 \\
8  & 42 & 0.826 & 0.815 & 0.846 & 0.786 \\
\textbf{10} & \textbf{40} & \textbf{0.836} & \textbf{0.821} & \textbf{0.842} & \textbf{0.800} \\
12 & 36 & 0.805 & 0.773 & 0.744 & 0.806 \\
15 & 33 & 0.835 & 0.794 & 0.771 & 0.818 \\
\hline
\end{tabular}
\end{table}

\textcolor{red}{For each element, we compared the concentration distributions of correctly versus incorrectly classified alloys (\textbf{Table} \ref{tab:element_permutation}) using a two-sided permutation test of the difference in means with 10,000 resamples. We selected a permutation test because it rests on minimal assumptions: under the null hypothesis of no difference in mean concentration between correctly and incorrectly classified alloys, the group labels are exchangeable, so the null distribution is built by relabeling the data rather than by assuming an underlying distribution \cite{ernst2004permutation}. This suits the small, bounded, and non-normal compositional distributions in this dataset, for which the normality and large-sample approximations of parametric tests are unreliable. We corrected for multiple comparisons across elements using the Benjamini-Hochberg false discovery rate procedure \cite{benjamini1995controlling}. No element showed a significant difference between correctly and incorrectly classified alloys, either before or after correction (smallest uncorrected p = 0.057 for Hf, smallest FDR-adjusted p = 0.48). We note explicitly that, given the small number of misclassified alloys (n = 4--14 per element among testable elements), this test has limited statistical power and can detect only large effect sizes. The non-significant result therefore indicates the absence of a large single-element compositional signature rather than evidence that no difference exists \cite{altman1995statistics}. Several elements present in too few misclassified alloys (2 or fewer misses) could not be tested and are reported as such.}

\begin{table}[H]
\centering
\caption{Per-element comparison of mean concentration between correctly (hit) and incorrectly (miss) classified alloys, restricted to alloys in which each element is present. $p$-values are from two-sided permutation tests (10{,}000 resamples); $p_\mathrm{FDR}$ is the Benjamini--Hochberg adjusted value across tested elements. Elements with $n_\mathrm{miss} \leq 2$ were not tested (---).}
\label{tab:element_permutation}
\begin{tabular}{lcccccc}
\hline
Element & $p$ & Hit mean & Miss mean & $n_\mathrm{hit}$ & $n_\mathrm{miss}$ & $p_\mathrm{FDR}$ \\
\hline
Hf & 0.057 & 0.231 & 0.173 & 23 & 4  & 0.482 \\
Ta & 0.107 & 0.205 & 0.247 & 36 & 7  & 0.482 \\
Mo & 0.223 & 0.176 & 0.224 & 44 & 6  & 0.670 \\
Cr & 0.540 & 0.227 & 0.237 & 19 & 4  & 0.968 \\
Nb & 0.543 & 0.224 & 0.236 & 33 & 7  & 0.968 \\
Ni & 0.723 & 0.272 & 0.283 & 23 & 4  & 0.968 \\
W  & 0.894 & 0.188 & 0.185 & 73 & 14 & 0.968 \\
Re & 0.925 & 0.229 & 0.232 & 18 & 4  & 0.968 \\
V  & 0.968 & 0.168 & 0.165 & 18 & 4  & 0.968 \\
\hline
Ti & --- & 0.141 & 0.200 & 16 & 1 & --- \\
Al & --- & 0.016 & 0.054 & 3  & 1 & --- \\
Co & --- & 0.267 & 0.290 & 14 & 2 & --- \\
Fe & --- & 0.230 & 0.224 & 14 & 2 & --- \\
Zr & --- & 0.144 & 0.258 & 7  & 2 & --- \\
Ru & --- & 0.230 & 0.267 & 15 & 2 & --- \\
\hline
\end{tabular}
\end{table}
\textcolor{red}{To determine what does characterize the misclassified alloys, we examined their position relative to the SVC decision boundary in \textbf{Table} \ref{tab:decision_function}. For a support vector classifier, the decision function returns a signed distance from each sample to the separating hyperplane in the model's feature space; its magnitude reflects the model's confidence, with values near zero indicating samples close to the boundary where the classification is most ambiguous. Comparing the absolute decision-function values of correctly and incorrectly classified alloys, misclassified alloys lie significantly closer to the decision boundary than correctly classified alloys (median |distance| 0.745 versus 1.233; one-sided permutation test, p = 0.0002). This indicates that the model's errors are concentrated among lower-confidence cases near the decision boundary rather than across a wider region of feature space, consistent with the absence of a per-element compositional signature reported above. Together, these analyses indicate that the model reliably classifies compositions that are well-separated in feature space and fails primarily on borderline cases, which is consistent with how margin-based classifiers are expected to behave.}
\begin{table}[htbp]
\centering
\caption{Absolute SVC decision-function values (distance to the decision boundary) for correctly (hit) and incorrectly (miss) classified alloys. Misclassified alloys lie significantly closer to the boundary (one-sided permutation test, $p = 0.0002$).}
\label{tab:decision_function}
\begin{tabular}{lccc}
\hline
 & Median $|d|$ & Mean $|d|$ & $n$ \\
\hline
Hit  & 1.233 & 1.549 & 73 \\
Miss & 0.745 & 0.734 & 14 \\
\hline
\end{tabular}
\end{table}
%%%%%%%%%%%%%%%%%%%%%%%%%%%%%%%%%%%%%%%%%%%%%%%%%%%%%%%%%%%%%%%%%%%%%%%%%%%%%%
\subsubsection{\textcolor{red}{Electronic Structure Calculations}}

\begin{figure}[H]
\centering
\includegraphics[width=\textwidth]{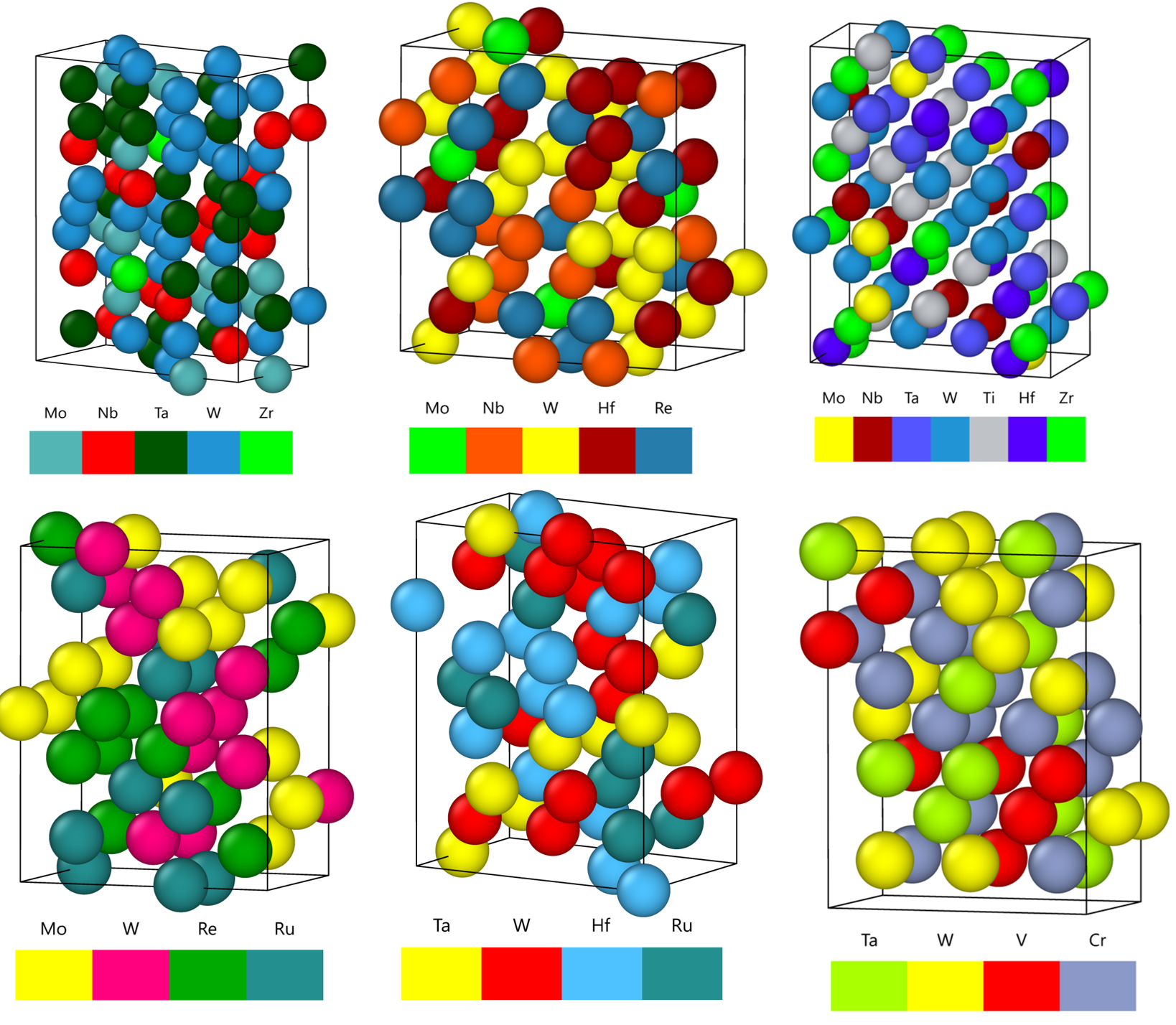}
\caption{\textcolor{red}{Special quasirandom structure (SQS) supercells used for first-principles calculations of the selected validation alloys. Each structure was generated using ICET from a bcc parent lattice and populated with the target elemental concentrations to reproduce the correlation functions of an ideal random solid solution. Different colors denote distinct chemical species. The displayed configurations correspond to the initial SQS structures prior to full DFT relaxation.}}
\label{fig:sqs_structures}
\end{figure}

{\color{red}
First-principles calculations were performed to validate the ductility classifications predicted by the SVC. From the ML-screened alloy space, a representative set of candidate alloys predicted to exhibit either ductile or brittle behavior was selected for detailed DFT calculations. Owing to the finite-size requirements of special quasirandom structure (SQS) supercells, the exact ML-predicted compositions were approximated by nearby commensurate compositions with integer atomic occupancies. These simplified compositions preserve the elemental chemistry and relative concentration trends of the original candidates while enabling the construction of chemically disordered supercells suitable for first-principles calculations. Chemical disorder was modeled using the SQS approach as implemented in the ICET package. For each selected alloy, a body-centered cubic (bcc) reference lattice was constructed using elemental W as the parent structure. The target alloy compositions were then mapped onto finite supercells by assigning chemical species to lattice sites such that the resulting pair and multisite correlation functions closely reproduced those of an ideal random solid solution. Supercell sizes between 48 and 80 atoms were employed depending on the number of constituent elements and the minimum elemental concentration. The final supercell size and elemental occupancies were chosen to provide the closest realizable approximation to the target composition while maintaining computational tractability. \textbf{Fig.}~\ref{fig:sqs_structures} shows the resulting SQS configurations used in the DFT calculations. Although the initial structures are generated from a common bcc parent lattice, the atomic occupations are optimized by the SQS procedure to emulate chemical randomness while preserving the target alloy composition. These structures serve only as the starting point for the first-principles calculations; all atomic positions and lattice vectors were subsequently relaxed within DFT, allowing the local atomic environments and equilibrium cell dimensions to evolve naturally according to the energetics of each alloy.
}

{\color{red}
All DFT calculations were performed using the Vienna \textit{Ab Initio} Simulation Package (VASP) \cite{kresse1993,kresse1996}. Electron--ion interactions were described using the projector augmented-wave (PAW) method \cite{blochl1994,kresse1999}, and exchange--correlation effects were treated within the generalized gradient approximation using the Perdew--Burke--Ernzerhof (PBE) functional \cite{pbe1996}. Spin-polarized calculations were performed for all compositions. Plane-wave basis functions were expanded to an energy cutoff of 550 eV, and $\Gamma$-centered Monkhorst--Pack $k$-point meshes were generated according to the dimensions of each SQS supercell. Each SQS structure was fully relaxed with respect to both atomic positions and lattice vectors until the total energy and residual atomic forces were converged. Consequently, while the ICET-generated structures provide the chemically disordered starting configurations, all reported electronic and mechanical properties were obtained from fully relaxed DFT equilibrium structures. The relaxed structures were subsequently used in static self-consistent field (SCF) calculations to obtain accurate electronic energies, charge densities, and density-of-states (DOS) information. Static calculations employed an electronic energy convergence criterion of $10^{-7}$ eV and Gaussian smearing of 0.10 eV. To ensure a comprehensive and rigorous characterization of the six candidate alloys, a total of 168 independent DFT calculations were executed across this workflow, comprising 6 full structural relaxations, 6 subsequent static SCF runs, 150 finite-strain calculations (25 configurations per alloy) for elastic property derivation, and 6 specialized DOS evaluations.
}

{\color{red}
Elastic properties were evaluated using a finite-strain stress--strain approach. Starting from the fully relaxed structure, six independent deformation modes corresponding to the Voigt strain components were applied. For each strain mode, positive and negative strains of $\pm0.5\%$ and $\pm1.0\%$ were imposed, resulting in a total of twenty-five calculations per alloy, including one unstrained reference configuration and twenty-four strained configurations. The stress tensors obtained from the converged SCF calculations were used to reconstruct the full elastic stiffness tensor $C_{ij}$ through a linear stress--strain relationship. Because the SQS structures possess no ideal crystallographic symmetry, no symmetry constraints were imposed during the fitting procedure. The calculated elastic stiffness tensor was used to determine the Voigt, Reuss, and Hill averaged bulk modulus ($B$) and shear modulus ($G$). Young's modulus ($E$), Poisson's ratio ($\nu$), and the Pugh ratio ($B/G$) were subsequently obtained from the Hill averages. The Vickers hardness was estimated from the shear modulus using the empirical relationship
\begin{equation}
H_V = 0.79 + 0.10G + 1.47\times10^{-4}G^2
\label{eq:hv}
\end{equation}
where $H_V$ and $G$ are expressed in GPa. Although this hardness estimate is derived from the elastic response and does not explicitly model plastic deformation, it provides a useful comparative descriptor of mechanical resistance across the investigated alloys. The DFT-derived mechanical descriptors, particularly the Pugh ratio, elastic moduli, and estimated hardness, were used as independent indicators of ductility and compared directly with the SVC classifications. Additional electronic-structure analysis was performed using total and projected density-of-states calculations to examine differences in metallic bonding, electronic delocalization, and bonding character between alloys predicted to be ductile and brittle.
}

%%%%%%%%%%%%%%%%%%%%%%%

\section{Results and Discussion}

\subsection{Model Results}

\begin{figure}[H]
    \centering
    \includegraphics[width=\textwidth]{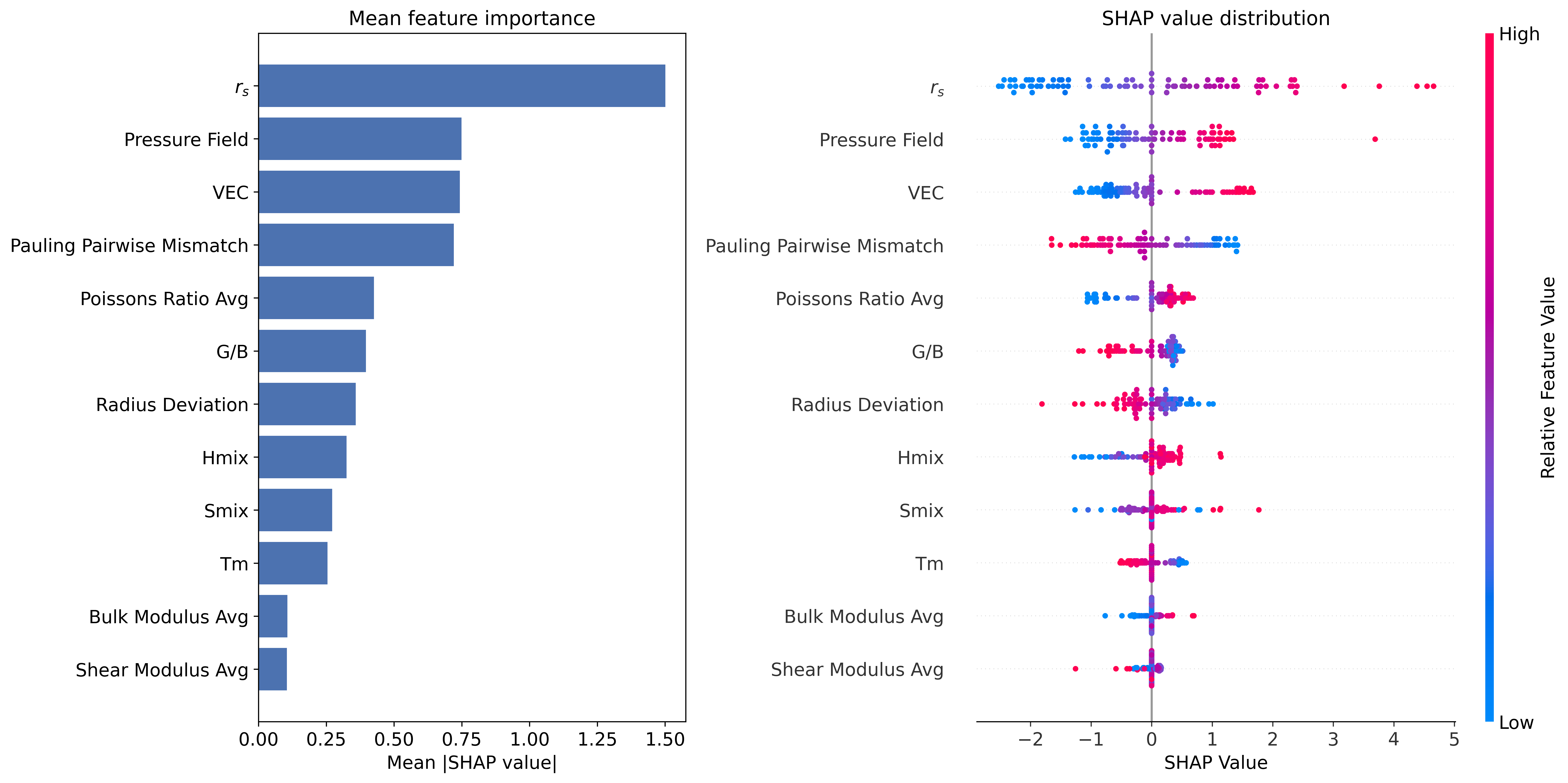}
    \caption{Shapley additive explanations (SHAP) summary of each feature's contribution to the support vector machine's output. Right: SHAP summary (beeswarm) plot, where point colors indicate low (blue) to high (red) feature values and position along the x-axis represents each feature's impact on the model output. Left: mean absolute SHAP value per feature, ranking overall feature importance.}
    \label{fig:SVC_shap}
\end{figure}

The SVC model was trained on the entire dataset using the same hyperparameter optimization strategy validated during model selection. To understand how the model formed its decision boundary, Shapley additve explanation (SHAP) values \cite{lundberg2017unified} were calculated for all input features, and the resulting summary plot is shown in \textbf{Fig.~\ref{fig:SVC_shap}}.

Shapley values are a game-theoretic approach to attribute the credit or blame for the outcome of a game to its players by examining how the outcome of the game changes with different combinations of players.\cite{shapley1953value} In the context of machine learning, the players become the features (inputs) of a model, and the outcome of the game becomes the prediction of the model \cite{lundberg2017unified}, with the Shapley values themselves being a measure of how much each feature's value for a given instance pushes the prediction toward or away from the base rate (average prediction).

For a given alloy instance $x$, the model output can be decomposed as
\begin{equation}
f(x) = \mathbb{E}[f(x)] + \sum_i \phi_i,
\end{equation}
where $\mathbb{E}[f(x)]$ represents the baseline value (the average model output over the training dataset) and $\phi_i$ is the SHAP value associated with feature $i$. A SHAP value of $0$ therefore indicates that the feature does not shift the prediction relative to the dataset average. Positive SHAP values move the prediction toward the ductile class (i.e., toward a positive decision function output), whereas negative SHAP values push the prediction toward the brittle class.

The feature value is simply the measured input descriptor for that instance (e.g., the actual VEC or $H_{\mathrm{mix}}$ of a given alloy). In the SHAP beeswarm plot (right side of \textbf{Fig.} \ref{fig:SVC_shap}), color encodes whether this value is high (red) or low (blue) relative to the range of values for that feature across all instances in the dataset. Each point in the beeswarm plot represents one alloy instance. Its horizontal position indicates the SHAP value (i.e., the magnitude and direction of that feature’s contribution to the prediction), while the color indicates whether that instance had a relatively high or low value for that feature. Features with a wider horizontal spread of SHAP values across instances have greater overall influence on the model output.

{\color{red}The accompanying bar plot (left side of \textbf{Fig.} \ref{fig:SVC_shap}) summarizes overall feature importance using the mean absolute SHAP value, computed for each feature as the average of $\lvert \phi_i\rvert$ across all instances, where $\phi_i$ is the SHAP value of that feature for instance $i$. This collapses the per-instance contributions shown as points in the beeswarm plot into a single magnitude per feature, quantifying how strongly each descriptor influences the model output on average regardless of the direction of that influence. Features with larger mean absolute SHAP values exert greater overall influence on the classification. The absolute value discards sign, so this ranking reflects the strength of a feature's contribution but not whether it pushes predictions toward the ductile or brittle class. The directional information is retained in the beeswarm plot. This analysis provided a quantitative ranking of the most influential descriptors contributing to the model's classification of ductile and brittle alloys. The SHAP dependence plots show how each descriptor's contribution varies across its range. Higher values of the exchange--correlation parameter are associated with positive SHAP contributions, shifting alloys toward the ductile side of the decision boundary, while lower values contribute negatively. Because the exchange--correlation parameter and valence electron concentration are strongly anti-correlated (Pearson $r = -0.82$), the same trend appears in VEC, and the two should be read as a single collinear signal rather than as independent effects. These transitions are gradual rather than sharp step functions, but they define quantitative regions in descriptor space that are associated with ductile versus brittle classifications. We interpret this as an empirical description of the learned decision surface rather than a causal physical mechanism or a single-element heuristic.}

\textcolor{red}{The SHAP analysis revealed that the four most important features combine electronic and structural aspects of bonding in refractory HEAs. These features were the exchange–correlation parameter $r_s$, the pressure field, the VEC, and the electronegativity mismatch. The exchange–correlation parameter $r_s$ is the density parameter of a homogeneous electron gas, defined through the interstitial electron density $\rho_0$}

\begin{equation}
\frac{4}{3}\pi r_s^{3} = \frac{1}{\rho_0}
\end{equation}

\textcolor{red}{so that smaller $r_s$ corresponds to higher interstitial electron density. Johnson et al \cite{johnson2023universal} showed that $\rho_0$ (equivalently $r_s$) is a universal physics-based metric for the maximum strength of metals and alloys: elastic moduli and maximum shear strength fall on a single electron-gas curve as a function of $r_s$, with a smaller $r_s$ indicating higher strength. Importantly, they demonstrate that a ROM estimate of $r_s$ from the constituent elements reproduces the directly computed DFT value, so that even as a simple compositional feature $r_s$ retains a meaningful connection to the underlying electronic structure. We therefore include the ROM $r_s$ as a feature because it provides an inexpensive descriptor that encodes electron-density and bonding information not captured by purely geometric or thermodynamic descriptors, rather than as a direct measure of ductility. We note that in the original work $r_s$ is a strength metric, and ductility is assessed separately through a Pugh-ratio estimate. The association  between $r_s$ and ductile classification as reported here is therefore an empirical relationship learned by the model.}

\textcolor{red}{The valence electron concentration is the average number of valence electrons per atom, a measure of d-band filling, whereas $r_s$ reflects the strength of metallic bonding through the interstitial electron density. In our dataset they are strongly anti-correlated (Pearson $r=-0.82$), so the model cannot cleanly separate their individual contributions, and the per-feature SHAP directions and magnitudes for $r_s$ and VEC should be interpreted jointly rather than as independent effects. In particular, because high $r_s$ coincides with low VEC across our compositions, the apparent association of higher $r_s$ with ductility cannot be disentangled from the VEC trend, and we do not interpret the $r_s$ direction mechanistically in isolation. The robust observation is instead that both rank among the most important features. Electronic-structure information, even when estimated inexpensively from a rule-of-mixtures, is the primary driver of the model's predictions, rather than the geometric or thermodynamic descriptors.}

\begin{figure}[H]
    \centering
    \includegraphics[width=1\textwidth]{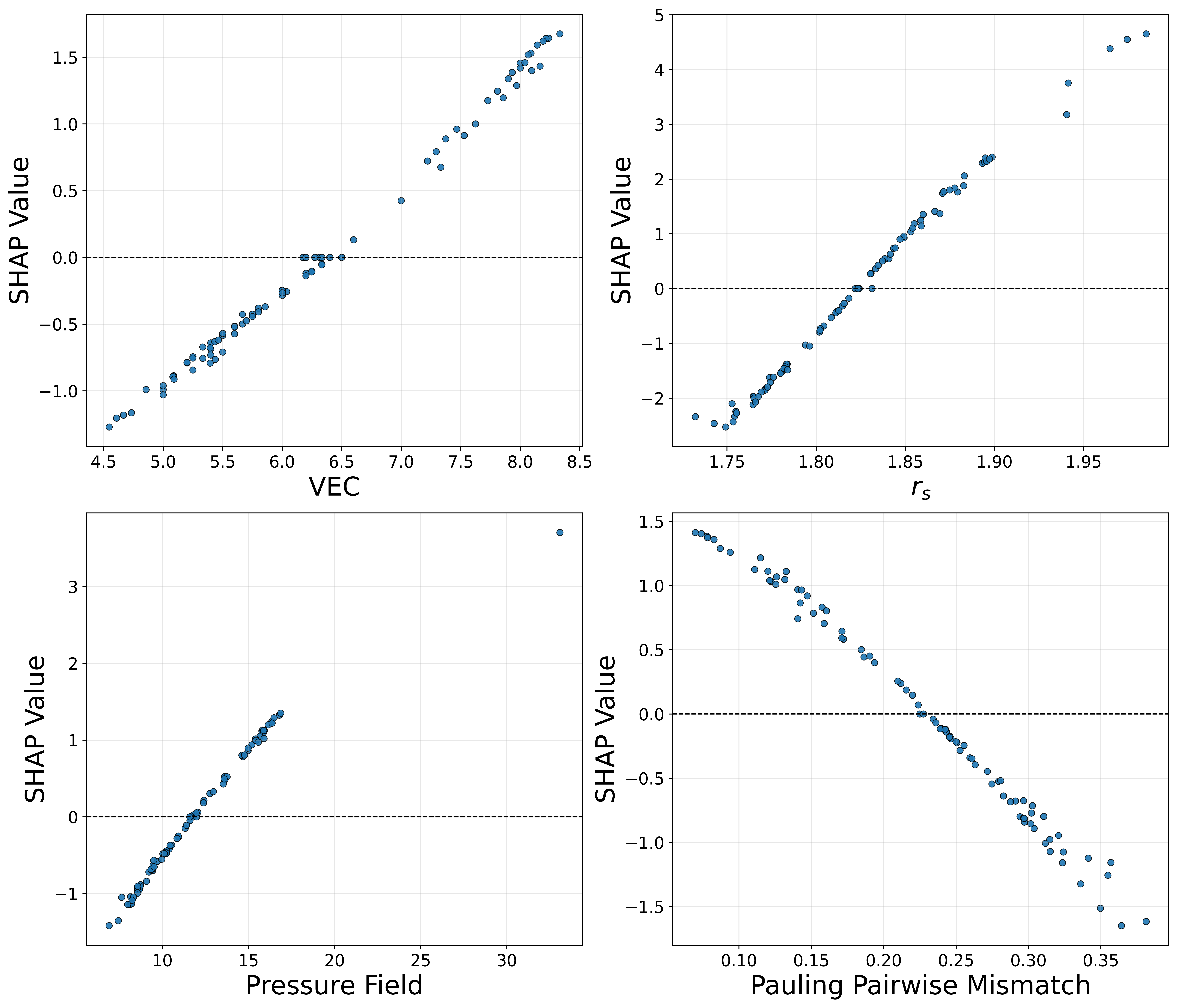}
    \caption{Dependence of Shapley additive explanations values on the four most influential features, illustrating how valence electron concentration, exchange correlation parameter, and pressure field promote ductile classifications while higher electronegativity mismatch favors brittle behavior. The sign of the Shapley additive explanations values is indicative of the direction of influence on model decision, with a value of 0 corresponding to no influence.}
    \label{fig:shap_dependence}
\end{figure}

\textcolor{red}{The pressure field descriptor originates from solid-solution-strengthening theory \cite{varvenne2016theory}. It is the hydrostatic stress (pressure) field of a gliding dislocation, which couples to the misfit volumes of the surrounding solute atoms. This elastic size-misfit interaction is the dominant contribution to solute strengthening in concentrated alloys, and as a compositional feature the pressure field scales with the shear modulus $\mu$ and Poisson's ratio $\nu$ as $\mu(1+\nu)/(1-\nu)$, thereby encoding how alloying modifies the elastic stiffness experienced by moving dislocations. As with $r_s$, it is fundamentally a strength-related quantity, so its association with the ductile class is an empirical relationship learned by the model rather than a direct ductility mechanism. The electronegativity mismatch quantifies chemical dissimilarity among the constituent elements and serves as an approximate measure of local bonding heterogeneity, which may promote or suppress cohesion depending on composition. Although a simplified representation of the complex interactions in multicomponent alloys, it may capture aspects of cohesion and structural stability relevant to fracture resistance. Together, these descriptors provide electronically and elastically motivated compositional features, and the dependence of the Shapley values on each is shown in \textbf{Fig.}~\ref{fig:shap_dependence}.}

\textcolor{red}{The partial-dependence trends indicate how these rule-of-mixtures parameters relate to predicted ductility in the W-containing HEAs of our dataset. Increasing VEC, particularly above approximately 6.5, generally shifts predictions toward the ductile class. The reported relationship between VEC and ductility is directional but not uniform. For many refractory BCC alloys, lower VEC is associated with greater ductility \cite{qi2014tuning, yang2021correlation}. In other cases,  higher VEC enhances ductility. For example, Re raises VEC yet is an effective ductilizer of W \cite{shaikh2023designing}, and in W-containing B1 carbides higher VEC promotes fracture resistance through electronic rehybridization and transformation-induced plasticity \cite{sangiovanni2023}. These behaviors arise from different mechanisms. Because our dataset spans varied W-HEA compositions, including Re-containing alloys, we interpret the model's VEC trend as an empirical relationship that captures part of this varied behavior rather than a single causal law. This is more consistent with the broader finding that electronic-structure governs the mechanical behavior of these alloys.}

\textcolor{red}{The pressure field showed a positive contribution to ductile classification. Because the pressure field is a strength-related, solute-strengthening quantity, we do not assign it a mechanistic ductility interpretation and treat this direction as an empirical association. The exchange--correlation parameter $r_s$ also showed a positive association with ductile classification. Since lower $r_s$ corresponds to higher electron density and higher theoretical strength \cite{johnson2023universal}, and $r_s$ is strongly anti-correlated with VEC in our dataset, this direction is consistent with the low-VEC/high-$r_s$ tendency toward ductility discussed above, although the collinearity prevents an independent causal interpretation. A lower electronegativity mismatch promoted ductile predictions, consistent with reduced bond heterogeneity and the improved solid solubility expected for more chemically compatible elements, which tends to suppress brittle intermetallic formation. This descriptor is a simplified representation of complex multicomponent interactions, but may capture aspects of cohesion and phase stability relevant to fracture resistance.}

%%%%%%%%%%%%%%%%%%%%%%%%%%%%%%%%%%%%%%%%%%%%%%%%%%%%%%%%%%

\subsection{Visualizing the Design Space}
\label{sec:visualizing_SVC}

Visualizing the design space of HEAs is challenging because of the inherently high dimensionality of multi-element composition spaces. A common approach when using ROM-based parameters is to employ pseudo-ternary phase diagrams \cite{ouyang2023design}. Although the features used for training were averaged quantities, the underlying data are fundamentally compositional. Therefore, to interpret alloying effects directly in elemental space, we generated a synthetic dataset of alloys from the experimental alloys. \textcolor{red}{Each composition was perturbed by multiplicative Gaussian noise  with standard deviation 0.3 (30\%) applied with the same scale to every element, followed by renormalization to unit composition. We chose this as the simplest way to sample around existing alloys for visualizing the decision boundary in elemental space and acknowledge it is a deliberately unsophisticated scheme.} This sampling procedure preserved the elemental combinations represented in the original dataset and produced a realistic mixture of ternary, quaternary, and quinary compositions that better resembled the chemical diversity of the training data. We further restricted our analysis to alloys with W fractions in the range $20 \leq c_W \leq 40$, ensuring that the visualized design space corresponded to the composition range of practical interest for W-containing RHEAs.

For each synthetic composition, the reduced set of ROM features (as determined by PCC pruning) was recalculated, and only those points falling within the minimum and maximum of each of the features of the training data in feature space were retained. This ensured that all model predictions represented interpolation rather than extrapolation, which is particularly important because machine learning models should not be used to generalize beyond their training domain. Each retained composition was then evaluated using the trained SVC model, and the signed distance from the decision boundary was used as a proxy of predicted ductility (positive values indicating ductile, negative indicating brittle, and distance indicating stability of prediction under new training data).

\begin{figure}[H]
    \centering
    \includegraphics[width=1\textwidth]{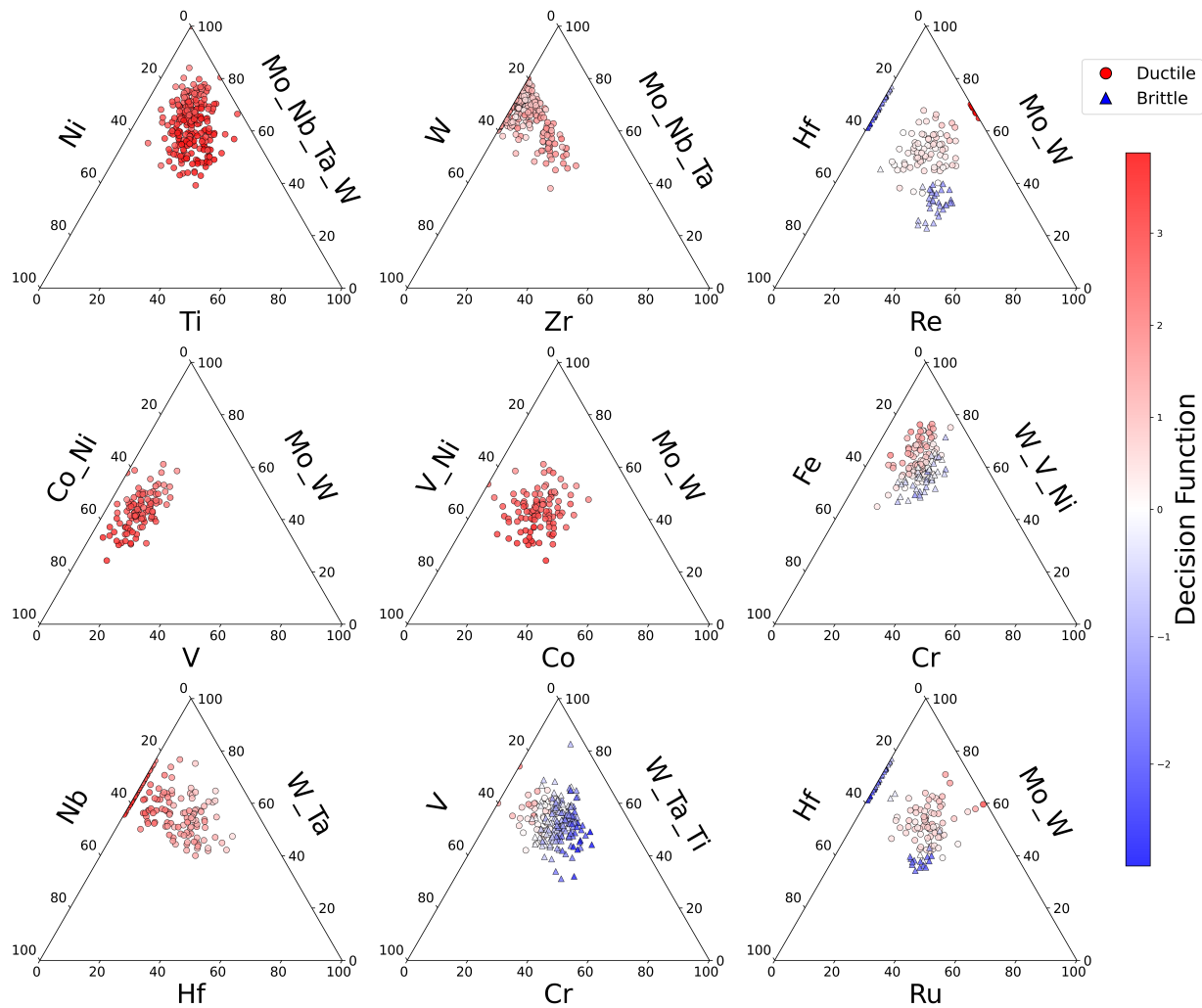}
    \caption{SVC decision function visualized across pseudo-ternary composition spaces for common alloying combinations in W-based RHEAs. Each point represents an interpolated synthetic composition colored by the signed decision value, where positive values in red indicate a ductile classification and negative values in blue correspond to brittle predictions. Additionally, points that are classified as ductile are represented by circles, and those classified as brittle are represented by triangles.}
    \label{fig:ternary}
\end{figure}

\textbf{Fig.~\ref{fig:ternary}} displays these decision-function values on pseudo-ternary diagrams for the most frequently occurring alloying combinations in the dataset. These maps reveal clear and physically meaningful alloying trends. In several systems, such as the (Ti, Ni, Mo--Nb--Ta--W) ternary, increasing Ti or Ni content consistently shifts compositions toward the ductile side of the decision function, forming broad regions of positive predictions. By contrast, ternaries containing Cr, such as (Cr, Fe, W--V--Ni), show large regions with negative decision values, indicating strong model confidence toward brittle behavior. These examples demonstrate that the continuous decision-function projections capture interpretable relationships between elemental composition and predicted ductility in W-containing HEAs.

It is also useful to examine how the SVC model partitions each pseudo-ternary space into discrete ductile and brittle domains. \textbf{Fig.} \ref{fig:ternary} shows the corresponding binary classification results obtained by thresholding the learned decision function. In Ti- and Ni-containing ternaries, the ductile region is extensive and contiguous, reflecting the stabilizing influence of these elements on the predicted mechanical response. In contrast, Cr-containing systems exhibit large, sharply bounded brittle regions, underscoring their tendency to suppress ductility in the present design space. The binary maps therefore complement the continuous decision function visualization by clarifying where transitions between ductile and brittle behavior occur.

Across these pseudo-ternary diagrams, several broad compositional trends emerge. Increasing the atomic fraction of Ti or Ni reliably raises the likelihood of ductility, especially in systems based on refractory elements such as W, Ta, Mo, and Nb. Cr is consistently associated with brittle predictions in W--V or W--Ni systems, reflecting its strong negative influence within the sampled design space. Ni and Co tend to produce ductile predictions most clearly when alloyed together. 

\begin{figure}[H]
    \centering
    \includegraphics[width=1\textwidth]{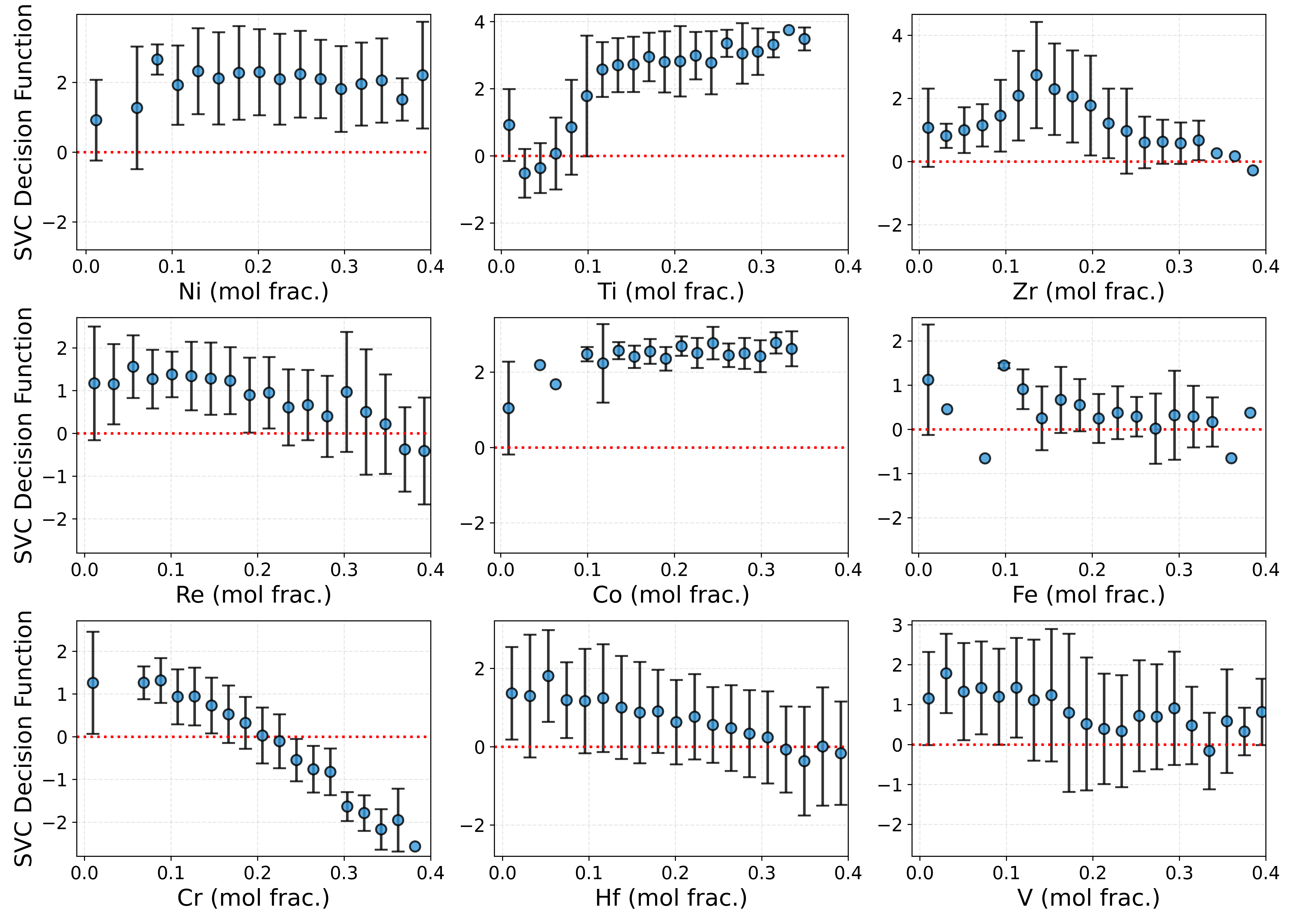}
    \caption{Element-wise dependence of the SVC decision function on composition for selected alloying elements. Synthetic compositions were binned by atomic fraction, and the points show the mean value of the decision function within each bin, with error bars indicating one standard deviation. The horizontal dashed line marks the learned boundary between ductile (positive) and brittle (negative) predictions.}
    \label{fig:strategy_grid}
\end{figure}

To further quantify these trends on an element-by-element basis, we visualized the dependence of the SVC decision function on the atomic fraction of individual alloying elements using binned compositional bar charts, which is a useful way to examine compositional effects on a property of a high-entropy alloy \cite{vela2025visualizing}. For each of nine selected elements, the synthetic dataset was binned by atomic fraction, and the mean and standard deviation of the decision function were plotted, as shown in \textbf{Fig.~\ref{fig:strategy_grid}}. A horizontal line at zero denotes the model’s decision boundary: compositions above this line are classified as ductile, and those below as brittle, with increasing distance indicating increasing prediction confidence. This analysis highlights which elements most strongly correlate with improved ductility within the interpolation space allowed by the dataset.

%%%%%%%%%%%%%%%%%%%%%%%%%%%%%%%%%%%%%%%%%%%%%%%%%%%%%%%%%%%%%%%%%%%%%%%%%%%%%%%%%%%%%%%%%%%
\subsection{Interpretation of Decision Boundary}

The pseudo ternary map in \textbf{Fig.~\ref{fig:ternary}}, together with the one dimensional compositional trends in \textbf{Fig.~\ref{fig:strategy_grid}}, suggest that the most reliable way to achieve room temperature ductility in W containing RHEAs is through targeted additions of Ti, Ni, and Co along with careful control of Cr contents. Among all alloying elements sampled, Ti produced the strongest and most consistent positive shift in the SVC decision function, indicating that increasing Ti content within the studied range systematically moves compositions toward the ductile side of the decision boundary. This result is consistent with experimental work on NbMoTaW and VNbMoTaW based RHEAs, where Ti additions have been shown to stabilize deformable BCC or BCC B2 microstructures, reduce the shear modulus, and markedly improve room temperature tensile ductility \cite{han2017effect,han2018microstructures,tsuru2024intrinsic}. Similar Ti driven improvements in ductility have also been reported in other refractory alloy systems such as HfNbTaTiZr and related TiZrNbHfTa V families, which reinforces the view that Ti is a broadly effective ductilizing element in BCC refractory alloys \cite{tsuru2024intrinsic, bai2021titanium, wang2024effect}.

The behavior of Ni and Co follows the same general pattern. \textbf{Fig.~\ref{fig:strategy_grid}} shows that higher Ni content, especially when combined with Co, consistently increases the predicted likelihood of ductility. This trend is aligned with the long standing use of Ni rich phases to impart toughness in classical tungsten heavy alloys and with more recent studies showing that Ni containing high entropy alloys achieve excellent combinations of strength and ductility through multiple deformation mechanisms. In the present compositional space, Ni and Co do not act independently but instead reinforce each other, and their combined presence broadens the region of compositions predicted to remain ductile.

In contrast, the trends for Cr indicate that an increasing atomic fractions typically reduces the likelihood of ductility. Systems containing Cr show narrow regions of positive decision function values and broad regions where the model predicts brittle behavior, which is consistent with the formation of complex topologically close packed phases reported in Cr rich refractory alloys. The same pattern appears in systems with significant Hf content. Although Hf can increase strength in certain alloy families, high Hf levels often promote multiphase microstructures with limited plasticity at room temperature. The model therefore reflects known challenges associated with Cr and Hf additions when the objective is to reduce the ductile to brittle transition temperature. Together these results indicate that the design space for ductile W based RHEAs is best explored through the controlled use of Ti, Ni, and Co, while avoiding excessive levels of Cr and Hf. The agreement between these model driven trends and the broader experimental literature suggests that the chosen classifier and validation structure captures meaningful physical relationships between elemental chemistry and mechanical behavior in our dataset.

The trends for Ni and Co in \textbf{Fig.~\ref{fig:strategy_grid}} likewise show that higher Ni contents, especially when combined with Co, increase the distance of synthetic compositions into the ductile regime. This behavior mirrors the long-standing use of Ni-rich binder phases to impart toughness in conventional W--Ni--Fe and W--Ni--Co heavy alloys, as well as more recent reports of Ni-rich high-entropy alloys that achieve exceptional strength--ductility synergy through complex dislocation and twinning mechanisms.\cite{jiang2015microstructure, nutor2021dual, li2024high} In the present W-based design space, Ni and Co do not act as separate ductile and brittle elements but rather as a cooperative pair, with Co helping to sustain ductility in Ni-containing compositions that would otherwise be too refractory-rich.

By contrast, the Cr and Hf panels in \textbf{Fig.~\ref{fig:strategy_grid}} exhibit monotonic trends toward more negative decision-function values as their atomic fractions increase, indicating an increased likelihood of brittle classification at higher concentrations. This is qualitatively consistent with reports that Cr-rich refractory HEAs tend to form topologically close-packed or Laves-type secondary phases that raise hardness and strength at the expense of ductility,\cite{yurchenko2024strength, toda2020effect} and with observations that Hf-rich refractory systems often develop complex multiphase microstructures that elevate the ductile-to-brittle transition temperature even when overall strength remains high \cite{badica2025high, vlasak2022thermal}. Overall, these results suggest that, within the W-containing RHEA space sampled here, alloying strategies that combine moderate Ti, Ni, and Co additions with limited Cr and Hf contents are most likely to yield alloys that remain ductile at room temperature while retaining the high-temperature capability associated with refractory elements. 

\textcolor{red}{
While the compositional trends identified here favor room-temperature ductility, deployment in fusion environments ultimately requires consideration of neutron irradiation effects. Irradiation introduces vacancy clusters, dislocation loops, and transmutation products that increase the ductile-to-brittle transition temperature and modify mechanical response \cite{Hu2016IrrHardening}. Thermal conductivity degradation above approximately 0.2 dpa has been reported in pure tungsten, with reductions approaching one-third of the unirradiated value \cite{Cui2018ThermalCond}. In addition, transmutation-induced rhenium and osmium accumulation contributes to resistivity changes beyond those caused by lattice defects alone \cite{Lang2023Transmutation}. In the present work, density functional theory calculations were performed to validate the elastic and electronic characteristics of the machine-learning-selected candidate alloys. These calculations provide independent first-principles support for the predicted ductility trends, but they do not explicitly account for irradiation-induced defect evolution, transmutation kinetics, or radiation-enhanced segregation. Consequently, the machine learning and DFT results should be interpreted as identifying ductility-favored compositional regions within the experimentally sampled space under unirradiated conditions. Extension of the present framework to include irradiation evolution, activation constraints, and thermal transport degradation represents an important direction for future work.}

\subsubsection{\textcolor{red}{Selection of Alloys for DFT Validation}}

On the basis of the continuous decision-function maps shown in \textbf{Fig.~\ref{fig:ternary}}, it is possible to identify specific compositions within the interpolation domain of the synthetic dataset that lie in strongly positive, strongly negative, and near-boundary decision-margin regions of the trained SVC model for comparison with DFT-derived features relevant to ductility. For each synthetic composition satisfying (i) $\sum_i c_i = 1$, and (ii) $20 \leq c_W \leq 40$, the reduced (PCC-pruned) set of ROM descriptors listed in \textbf{Table\ref{tab:parameters}} was computed. The resulting feature vector $\mathbf{x}$ was standardized using the training-set mean $\boldsymbol{\mu}$ and standard deviation $\boldsymbol{\sigma}$ as $\hat{\mathbf{x}} = (\mathbf{x}-\boldsymbol{\mu})/\boldsymbol{\sigma}$ prior to evaluation by the trained support vector classifier. The signed distance from the decision boundary was then obtained from the SVC decision function
\[
f(\hat{\mathbf{x}}) = \sum_{i=1}^{N_{sv}} \alpha_i y_i K(\hat{\mathbf{x}}_{sv,i}, \hat{\mathbf{x}}) + b,
\]
where $K$ is the radial basis function kernel, $\alpha_i$ are the learned dual coefficients, and $b$ is the intercept. {\color{red}To further interpret the learned SVC decision boundary and provide a physics-based validation of the machine-learning results, six representative alloys from the synthetic dataset were selected for first-principles calculations. The selected set spans both experimentally ductile and brittle compositions and includes alloys positioned on either side of the classifier boundary. Importantly, the compositions used here are not equiatomic; instead, they are the exact target fractional compositions carried forward into the SQS construction and subsequent DFT workflow.

For each candidate we found the closest training alloy made of the same elements and measured how far apart the two compositions are. Each alloy is a set of atomic fractions that sum to one. The distance $\Delta$ is the fraction of atoms that differ in identity between the two alloys: we sum the absolute per-element differences in atomic fraction and divide by two,
\begin{equation}
\Delta = \frac{1}{2} \sum_i \left| x_i^{\mathrm{cand}} - x_i^{\mathrm{train}} \right|,
\end{equation}
where $x_i$ is the atomic fraction of element $i$. The division by two reflects that any fraction added to one element must be removed from another, so the raw sum counts each change twice. A $\Delta$ of 0~at.\% means identical compositions, and larger values mean greater separation. For example, a candidate enriched in one element by 4~at.\% relative to an otherwise-equiatomic training alloy has $\Delta = 4$~at.\%. \textbf{Table}~\ref{tab:dft_candidates} shows that the selected candidates, while constrained to lie within the bounds of the training feature space, remain compositionally distinct from individual training alloys, with separations of 4.2--16.6~at.\%.

\textbf{Table}~\ref{tab:dft_candidates} summarizes the selected alloys together with their closest training set compositions, distance from the training set alloy, SVC prediction, and SVC decision score. The selected compositions cover a wide range of predicted mechanical response, from strongly ductile candidates to brittle candidates, allowing a direct comparison between the SVC classification and the DFT-derived mechanical descriptors. These target compositions were then mapped onto commensurate SQS supercells with integer atomic occupancies for the DFT calculations described below.
}

\begin{table}[H]
\centering
\small
\renewcommand{\arraystretch}{1.2}
\caption{\textcolor{red}{Selected non-equiatomic alloy compositions used for DFT validation, together with the nearest training alloy sharing the same element set. Distance ($\Delta$) is the total compositional change (sum of absolute atomic-fraction differences, in at.\%). The SVC prediction and score are for the DFT composition. The compositions shown represent the exact whole-number atomic counts used in the SQS supercells for DFT calculations.}}
\label{tab:dft_candidates}
\begin{tabular}{p{5.6cm} p{2.8cm} c c c}
\hline
\textbf{Exact DFT Composition} & \textbf{Nearest training alloy} & \textbf{$\Delta$ (at.\%)} & \textbf{SVC prediction} & \textbf{SVC score} \\
\hline
Mo$_{5}$Nb$_{7}$Ta$_{14}$W$_{17}$Ti$_{13}$Hf$_{9}$Zr$_{15}$ & HfMoNbTaTiWZr & 16.6 & Ductile & +4.37 \\
\hline
Mo$_{14}$W$_{12}$Re$_{11}$Ru$_{11}$ & MoReRuW & 4.2 & Ductile & +2.10 \\
\hline
Mo$_{4}$Nb$_{12}$W$_{20}$Re$_{13}$Hf$_{15}$ & HfMoNbReW & 15.0 & Ductile & +1.39 \\
\hline
Mo$_{12}$Nb$_{14}$Ta$_{23}$W$_{29}$Zr$_{2}$ & MoNbTaWZr$_{0.1}$ & 16.3 & Ductile & +0.79 \\
\hline
Ta$_{9}$W$_{15}$Hf$_{15}$Ru$_{9}$ & HfRuTaW & 12.5 & Ductile & +0.09 \\
\hline
Ta$_{10}$V$_{7}$W$_{14}$Cr$_{17}$ & CrTaVW & 14.6 & Brittle & -2.80 \\
\hline
\end{tabular}
\end{table}

{\color{red}
The exact compositions listed in \textbf{Table}~\ref{tab:dft_candidates} were subsequently used as the target compositions for the DFT study. Because these fractional compositions cannot always be represented exactly within finite supercells, special quasirandom structures (SQS) were constructed using commensurate integer occupancies that preserve the elemental chemistry and relative concentration trends of the original ML-selected alloys. The resulting SQS models were then used for electronic-structure and elastic-property calculations, as described in the following subsection.
}

%%%%%%%%%%%%%%%%%%%%%%%%%%%%%%%%%%%%%%%%%%%%%%%%%%%%%%%%%%%%%%%%%%%%%%%%%%%%%%

\subsection{\textcolor{red}{DFT Validation of Machine Learning Predictions}}

\begin{figure}[H]
\centering
\includegraphics[width=\textwidth]{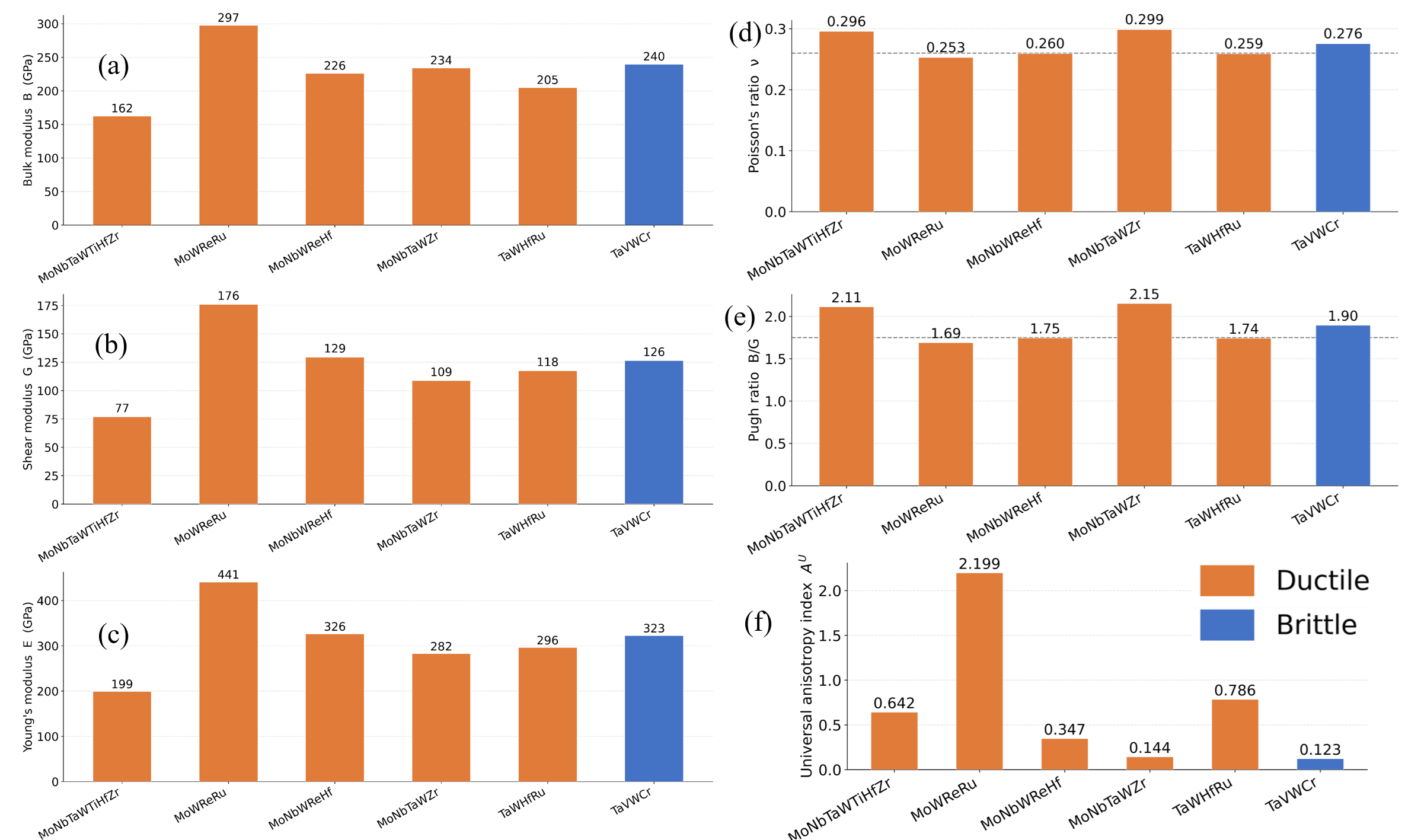}
\caption{
\textcolor{red}{
Density functional theory elastic validation of the machine learning selected refractory high entropy alloys. (a) Bulk modulus ($B$). (b) Shear modulus ($G$). (c) Young's modulus ($E$). (d) Poisson's ratio ($\nu$). (e) Pugh ratio ($B/G$). (f) Universal elastic anisotropy index ($A^{U}$). The dashed lines in panels (d) and (e) indicate the commonly accepted ductility thresholds of $\nu = 0.26$ and $B/G = 1.75$, respectively. The highest confidence machine learning candidates, MoNbTaWTiHfZr and MoNbTaWZr, exhibit the largest Poisson ratios and Pugh ratios among the investigated alloys, indicating mechanically favorable conditions for ductile deformation. These first principles calculations provide independent physical support for the machine learning predictions and demonstrate that alloys selected using the descriptor based decision boundary occupy favorable regions of elastic property space.
}
}
\label{fig:dft_elastic_validation}
\end{figure}

\textcolor{red}{The support vector classifier identified the exchange correlation parameter ($r_s$), valence electron concentration, pressure field, and electronegativity mismatch as the dominant descriptors governing the ductile to brittle classification boundary. Although these descriptors are electronic in origin, they ultimately influence atomic bonding, lattice resistance, and the ease of plastic deformation. To determine whether the machine learning selected alloys occupy physically meaningful regions of composition space, density functional theory calculations were performed on 6 candidate alloys (\textbf{Table} \ref{tab:dft_candidates}) from the synthetic dataset created above. The objective of these calculations was not to reproduce the classifier output directly. Instead, the calculations were designed to provide an independent first principles assessment of whether the candidate alloys identified by the machine learning framework exhibit elastic characteristics commonly associated with enhanced ductility. Elastic properties were extracted from the full elastic stiffness tensor obtained from finite strain calculations on the relaxed special quasi random structures. The resulting bulk modulus ($B$), shear modulus ($G$), Young's modulus ($E$), Poisson's ratio ($\nu$), Pugh ratio ($B/G$), and universal elastic anisotropy index ($A^{U}$) were subsequently used to evaluate the mechanical behavior of the machine learning selected compositions. The calculated elastic properties are summarized in \textbf{Fig.}~\ref{fig:dft_elastic_validation}. \textbf{Fig.}~\ref{fig:dft_elastic_validation}(a), (b), and (c) show the bulk modulus, shear modulus, and Young's modulus, respectively. The bulk modulus measures resistance to volumetric compression, whereas the shear modulus governs resistance to shape change and is more directly related to dislocation mediated deformation. Among the investigated alloys, MoNbTaWTiHfZr exhibits the lowest shear modulus of 77 GPa, while MoWReRu exhibits the highest shear modulus of 176 GPa. This difference indicates that MoNbTaWTiHfZr possesses a substantially lower resistance to shear deformation and therefore a greater tendency for plastic accommodation. More direct indicators of ductility are shown in \textbf{Fig.}~\ref{fig:dft_elastic_validation}(d) and (e), where Poisson's ratio and the Pugh ratio are compared with commonly accepted ductility criteria. The highest confidence machine learning candidates, MoNbTaWTiHfZr and MoNbTaWZr, exhibit the largest Poisson ratios of 0.296 and 0.299, respectively. The same alloys also exhibit the highest Pugh ratios of 2.11 and 2.15. Both quantities exceed the conventional ductility thresholds of $\nu = 0.26$ and $B/G = 1.75$, indicating favorable resistance to brittle fracture and enhanced capacity for plastic deformation. These observations provide independent support for the machine learning predictions and demonstrate that the highest confidence classifier candidates occupy mechanically favorable regions of elastic property space.}

\textcolor{red}{TaVWCr exhibits intermediate elastic properties and remains above the classical ductility thresholds. This observation indicates that isotropic elastic properties alone do not fully reproduce the machine learning decision function. Rather, the classifier appears to capture useful empirical relationships between ROM features and ductility that may not be represented by elastic constants alone. This result is consistent with the SHAP analysis, which identified $r_s$ and VEC as dominant electronic descriptors controlling the decision boundary. Both descriptors are fundamentally linked to electronic structure and bonding rather than purely macroscopic elasticity, which suggests that future study of W-HEAs should investigate their electronic structure more deeply. \textbf{Fig.}~\ref{fig:dft_elastic_validation}(f) presents the universal elastic anisotropy index. Lower values correspond to a more isotropic elastic response and generally indicate reduced directional dependence of deformation. MoNbTaWZr exhibits the smallest anisotropy index among the positive margin alloys, while MoWReRu exhibits the largest value. The combination of elevated Pugh ratio, high Poisson ratio, and relatively low anisotropy further supports the favorable mechanical behavior predicted for MoNbTaWZr by the machine learning model. Overall, the DFT calculations provide independent validation that the machine learning selected alloys exhibit elastic characteristics associated with enhanced ductility. While the elastic ranking does not exactly reproduce the classifier decision margins, the strongest ductile candidates identified by the support vector classifier consistently display the most favorable combinations of Poisson ratio, Pugh ratio, and shear resistance. These results indicate that the machine learning model captures physically meaningful relationships between composition, electronic descriptors, and mechanical response. To further examine the origin of these trends, the electronic structure of the candidate alloys is analyzed in the following section using density of states calculations.}

\subsubsection{\textcolor{red}{Hardness and Dislocation Energetics}}

\begin{figure}[H]
\centering
\includegraphics[width=\textwidth]{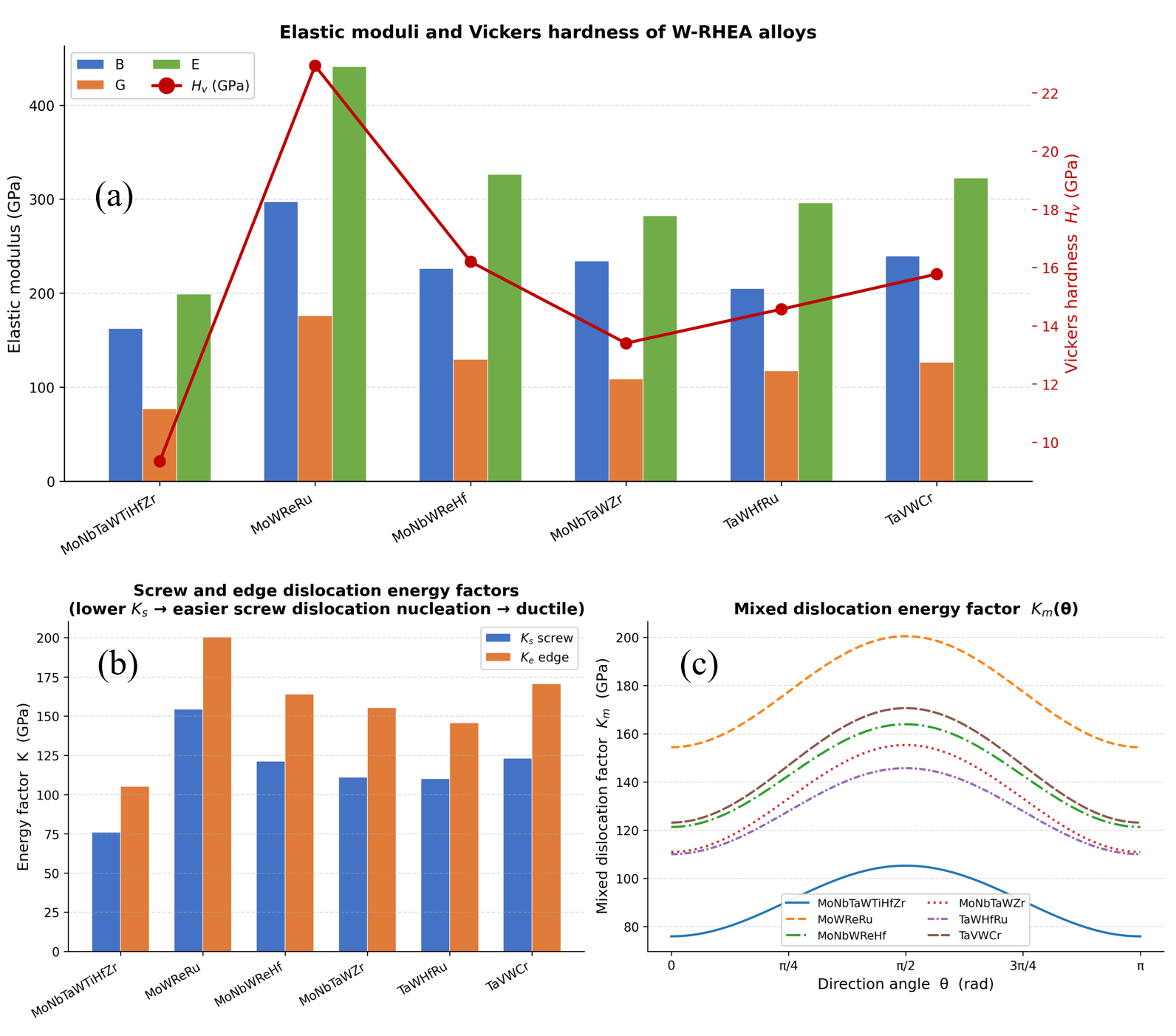}
\caption{
\textcolor{red}{
Density functional theory derived hardness and dislocation energetics of the machine learning selected refractory high entropy alloys. (a) Bulk modulus ($B$), shear modulus ($G$), Young's modulus ($E$), and estimated Vickers hardness ($H_V$). (b) Screw ($K_s$) and edge ($K_e$) dislocation energy factors derived from the elastic stiffness tensor. (c) Orientation dependent mixed dislocation energy factor ($K_m$). Lower values of the dislocation energy factors correspond to reduced energetic barriers for dislocation nucleation and motion.}
}
\label{fig:dft_dislocation_validation}
\end{figure}

\textcolor{red}{
The elastic analysis presented in \textbf{Fig.}~\ref{fig:dft_elastic_validation} demonstrated that the highest confidence machine learning candidates occupy mechanically favorable regions of elastic property space, exhibiting elevated Pugh ratios and Poisson ratios together with reduced resistance to shear deformation. However, the intrinsic brittleness of body centered cubic tungsten is fundamentally controlled by the nucleation and motion of screw dislocations rather than by isotropic elastic properties alone. Consequently, additional quantities were derived from the calculated elastic tensors to establish a more direct connection between the machine learning predictions and the underlying deformation mechanisms. \textbf{Fig.}~\ref{fig:dft_dislocation_validation}(a) summarizes the calculated elastic moduli together with the empirical Vickers hardness obtained from the methodology described in Section 2. The highest confidence ductile candidate identified by the support vector classifier, MoNbTaWTiHfZr, exhibits the lowest hardness among all investigated alloys, whereas MoWReRu exhibits the highest hardness. This behavior is consistent with the elastic trends observed in \textbf{Fig.}~\ref{fig:dft_elastic_validation}(b), where MoNbTaWTiHfZr displayed the lowest shear modulus and therefore the lowest resistance to plastic deformation. In contrast, the negative margin alloy TaVWCr exhibits a substantially larger hardness value, indicating a greater resistance to deformation despite possessing elastic properties that satisfy the classical ductility criteria.
}

\textcolor{red}{
The dislocation energy factors derived from the elastic stiffness tensors are presented in \textbf{Fig.}~\ref{fig:dft_dislocation_validation}(b) and \ref{fig:dft_dislocation_validation}(c). \textbf{Fig.}~\ref{fig:dft_dislocation_validation}(b) shows the screw and edge dislocation energy factors, while \textbf{Fig.}~\ref{fig:dft_dislocation_validation}(c) presents the orientation dependent mixed dislocation energy factor. The lowest screw dislocation energy factor is observed for MoNbTaWTiHfZr, followed by the remaining positive margin alloys, indicating a reduced energetic barrier for screw dislocation nucleation and propagation. Since DFT studies have established that the high ductile to brittle transition temperature of tungsten originates from the difficulty of activating and moving $\frac{1}{2}\langle111\rangle$ screw dislocations, a reduction in the screw dislocation energy factor is expected to promote dislocation mediated plasticity and improve ductility. The same alloy also exhibits the lowest mixed dislocation energy factor over the entire angular range considered. Importantly, these trends follow the ranking predicted by the support vector classifier. The alloy possessing the largest positive decision margin, MoNbTaWTiHfZr, consistently exhibits the lowest hardness, lowest screw dislocation energy factor, and lowest mixed dislocation energy factor among the investigated compositions. Likewise, MoNbTaWZr, which also lies well within the positive decision region of the classifier, exhibits comparatively favorable dislocation energetics. These observations provide a direct physical connection between the learned decision boundary and the atomistic deformation mechanisms governing tungsten based refractory high entropy alloys. The agreement between the classifier ranking and the calculated dislocation energetics therefore suggests that the machine learning framework successfully identifies compositions with intrinsically lower barriers to dislocation mediated plastic deformation. This result provides a mechanistic interpretation of the support vector classifier predictions and demonstrates that the learned descriptor space is physically consistent with the deformation physics known to govern ductility in tungsten based alloys.
}

%%%%%%%%%%%%%%%%%%%%%%%%%%%%%%%%%%%%%%%%%%%%%%%%%%%%%%%%%%%%%%%%%%%
\subsubsection{\textcolor{red}{Electronic Structure and Density of States}}

\begin{figure}[H]
\centering

\includegraphics[width=\textwidth]{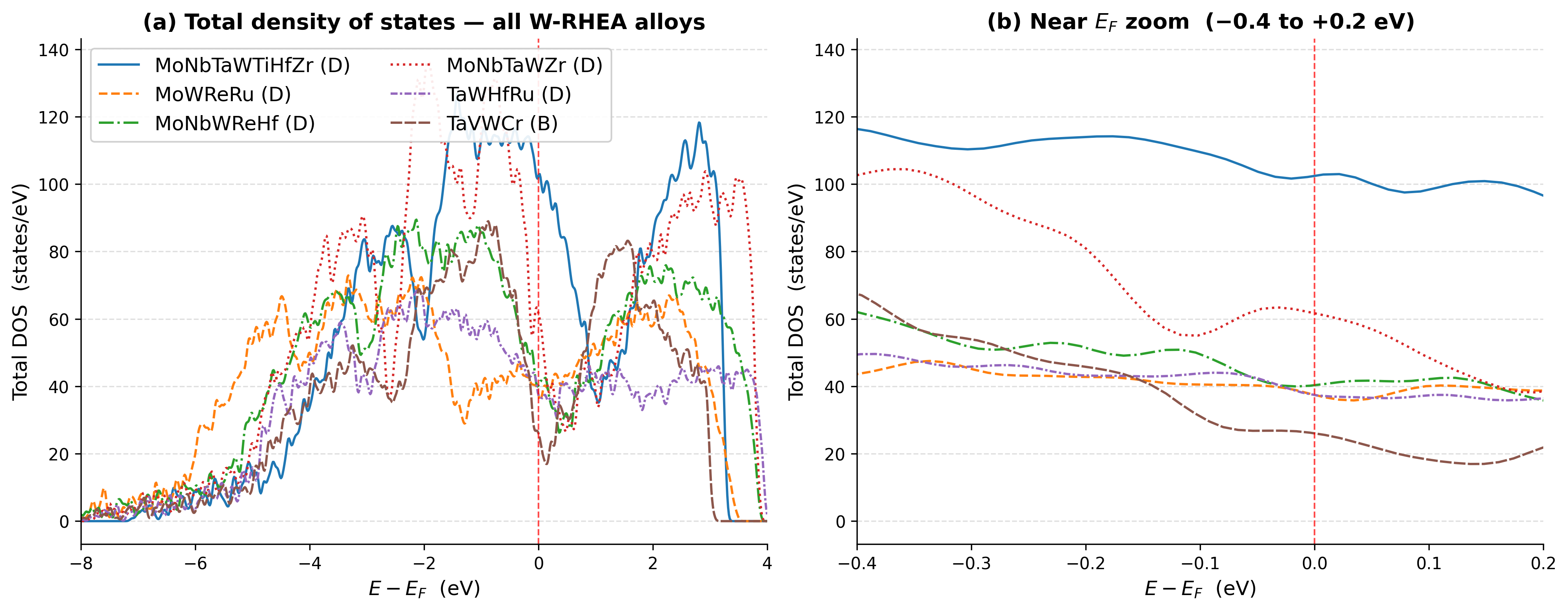}

\vspace{0.5cm}

\includegraphics[width=\textwidth]{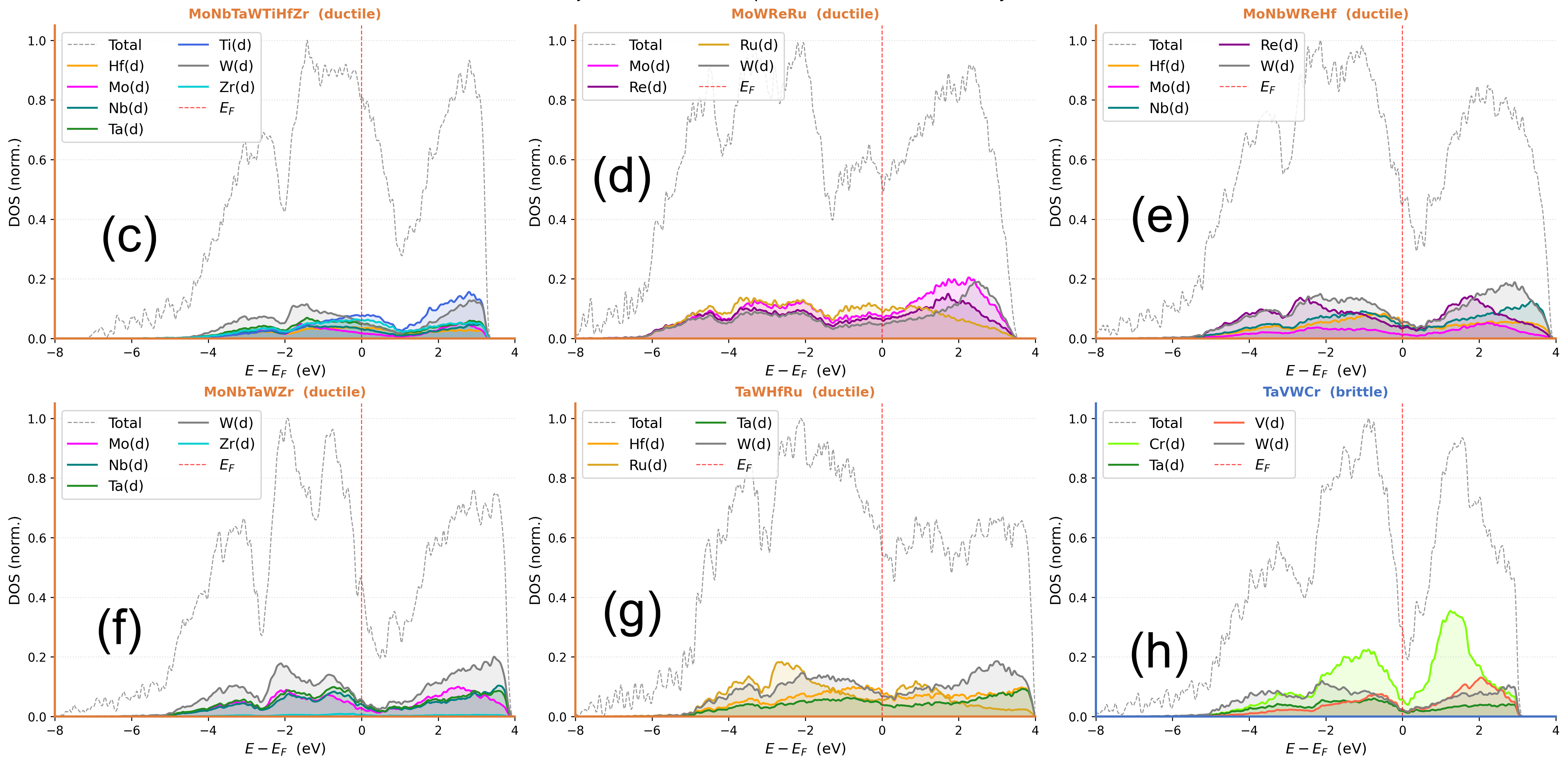}

\caption{
\textcolor{red}{
Density functional theory calculated electronic density of states (DOS) for the machine-learning-selected refractory high-entropy alloys. (a) Total DOS over the full energy range for all investigated compositions. (b) Enlarged view of the DOS near the Fermi level ($E_F$), where the vertical dashed line indicates the Fermi energy. (c) Atom-projected DOS of MoNbTaWTiHfZr. (d) Atom-projected DOS of MoWReRu. (e) Atom-projected DOS of MoNbWReHf. (f) Atom-projected DOS of MoNbTaWZr. (g) Atom-projected DOS of TaWHfRu. (h) Atom-projected DOS of TaVWCr. In panels (c)--(h), the gray dashed curve represents the normalized total DOS and the red dashed line denotes the Fermi level, $E_F$.
}}
\label{fig:dft_dos_validation}
\end{figure}

\textcolor{red}{
To further examine the electronic origin of the SVC predictions, the density of states (DOS) was calculated for all investigated alloys and is shown in \textbf{Fig.}~\ref{fig:dft_dos_validation}. The total DOS for the complete energy range is presented in \textbf{Fig.}~\ref{fig:dft_dos_validation}(a), while \textbf{Fig.}~\ref{fig:dft_dos_validation}(b) provides an enlarged view of the region surrounding the Fermi level, $E_F$. Since electronic states at or near $E_F$ participate most directly in bonding and deformation processes, the DOS at the Fermi level provides a useful indicator of the electronic environment experienced by moving dislocations. The highest-confidence ductile alloy, MoNbTaWTiHfZr, exhibits a broad and relatively smooth DOS near $E_F$, whereas the negative-margin alloy TaVWCr displays the most pronounced reduction in DOS around the Fermi level. The remaining positive-margin alloys occupy an intermediate regime. These differences indicate that the machine-learning-selected compositions possess distinct electronic structures despite exhibiting similar refractory BCC chemistries. Because VEC and the electronic descriptor $r_s$ were identified by the SHAP analysis as the dominant variables governing the SVC decision boundary, the DOS trends provide a direct electronic-structure perspective on the learned classifier behavior.
The atom-projected DOS shown in \textbf{Fig.}~\ref{fig:dft_dos_validation}(c)--(h) reveals the elemental contributions responsible for the total DOS behavior. In all alloys, states near $E_F$ are dominated by transition-metal $d$ orbitals, confirming that deformation-relevant bonding is controlled primarily by the refractory-metal $d$ electron network. For the positive-margin alloys, the contributions from the constituent elements are broadly distributed over energy, producing a more delocalized electronic structure around the Fermi level. In contrast, the brittle alloy TaVWCr, \textbf{Fig.}~\ref{fig:dft_dos_validation}(h), exhibits stronger localization of individual elemental $d$ states, particularly from Cr and V, within the vicinity of $E_F$. This distinction is consistent with the hardness and dislocation-energy trends discussed above. Together, the DOS results provide additional evidence that the SVC preferentially identifies compositions possessing electronic structures associated with lower lattice resistance and more favorable conditions for dislocation-mediated plastic deformation.
}

%%%%%%%%%%%%%%%%%%%%%%%%%%%%%%%%%%%%%%%%%%%%%%%%%%%%%%%%%%%%%%%%%%%%%%%%%%%%%%%%%

\subsubsection{\textcolor{red}{Relationship Between DFT-Derived Ductility Indicators and SVC Predictions}}

\begin{figure}[H]
\centering
\includegraphics[width=0.75\linewidth]{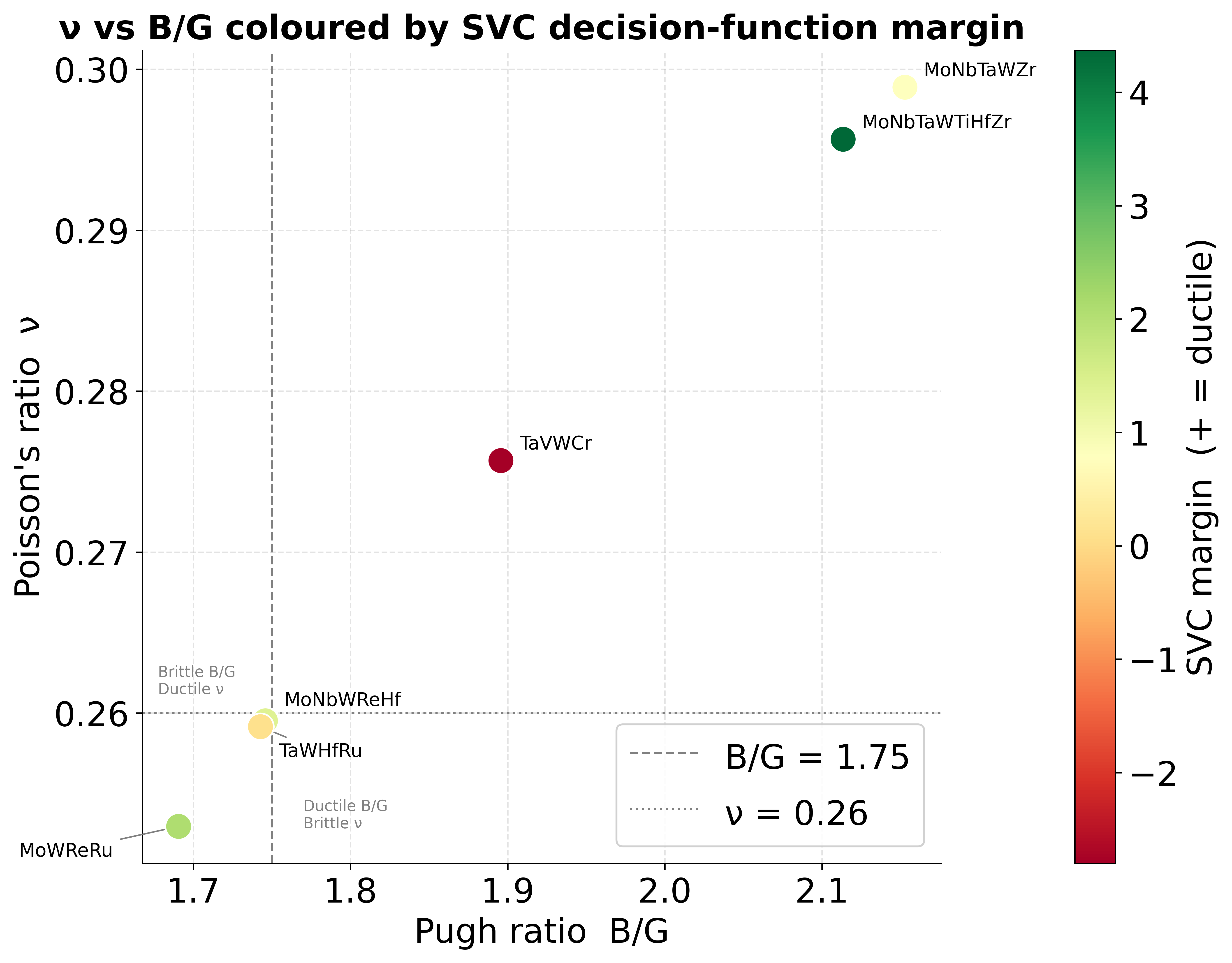}
\caption{\textcolor{red}{DFT-derived Poisson's ratio ($\nu$) plotted against the Pugh ratio ($B/G$) for the six alloys selected for first-principles validation. Data points are colored according to the SVC decision-function margin, where positive values indicate increasing confidence in the ductile class and negative values indicate increasing confidence in the brittle class. The dashed vertical line denotes the commonly used ductility criterion of $B/G = 1.75$, while the dotted horizontal line corresponds to $\nu = 0.26$.}}
\label{fig:dft_SVC_decision}
\end{figure}

{\color{red}
To examine whether the machine-learning predictions are consistent with independently calculated mechanical properties, the DFT-derived Pugh ratio ($B/G$) and Poisson's ratio ($\nu$) were compared with the SVC decision-function margins for the six validation alloys. These two elastic descriptors are among the most commonly used indicators of ductility in metallic systems and therefore provide a useful physics-based benchmark for evaluating the SVC predictions. \textbf{Fig.}~\ref{fig:dft_SVC_decision} presents the calculated values of $B/G$ and $\nu$ for each alloy, with the marker color indicating the corresponding SVC decision-function margin. Positive margins correspond to increasing confidence in the ductile class, whereas negative margins correspond to increasing confidence in the brittle class. The dashed vertical line and dotted horizontal line indicate the commonly used empirical ductility thresholds of $B/G = 1.75$ and $\nu = 0.26$, respectively. Several important trends emerge from \textbf{Fig.}~\ref{fig:dft_SVC_decision}. The alloy MoNbTaWTiHfZr, which received the largest positive SVC margin (+4.37), also exhibits one of the highest combinations of Pugh ratio and Poisson's ratio among the investigated compositions. Likewise, MoNbTaWZr occupies the upper-right region of the diagram and was assigned a positive decision score (+0.79). Both alloys satisfy the conventional ductility criteria and are clearly separated from the lower-ductility candidates.

At the opposite end of the decision boundary, TaVWCr received the only strongly negative SVC margin of the candidates ($-2.80$). Although its calculated Pugh ratio remains above the traditional threshold of 1.75, its comparatively lower Poisson's ratio and the fact that Cr is known experimentally to induce brittle behavior distinguish it from the more ductile compositions analyzed. This result illustrates an important limitation of relying on a single empirical criterion and highlights the need for multivariate approaches capable of incorporating multiple interacting descriptors simultaneously. The remaining alloys occupy intermediate positions in the descriptor space. In particular, MoWReRu, MoNbWReHf, and TaWHfRu exhibit relatively modest values of both $B/G$ and $\nu$ despite being assigned positive SVC margins. Notably, TaWHfRu lies very close to the classification boundary, with a decision score of only +0.09. Such alloys are inherently difficult to classify because small variations in composition, local chemical environment, experimental uncertainty, or model uncertainty can alter their predicted behavior. Similarly, MoNbWReHf and MoWReRu possess substantially smaller positive margins than MoNbTaWTiHfZr, indicating lower model confidence despite sharing the same predicted class.

Overall, \textbf{Fig.}~\ref{fig:dft_SVC_decision} demonstrates that the DFT-derived elastic descriptors broadly follow the trends learned by the SVC model. Alloys assigned large positive margins generally occupy regions associated with greater ductility, whereas alloys located near the classification boundary exhibit more ambiguous behavior. The agreement is not perfect, which is expected given that ductility is governed by a complex combination of elastic, electronic, and microstructural factors that extend beyond linear elastic response alone. Nevertheless, the consistency between the SVC decision scores and the independently calculated mechanical indicators suggests that the machine-learning framework captures physically meaningful trends and provides a useful predictive tool for identifying promising ductile refractory high-entropy alloys.
}

%%%%%%%%%%%%%%%%%%%%%%%%%%%%%%%%%%%%%%%%%%%%%%%%%%%%%%%%%%%%%%%%%%%%%%%%%%%%%%%%%%%

\section{Conclusion}

A curated dataset of tungsten containing refractory high entropy alloys was assembled from the literature, and composition based, physics informed descriptors were calculated as inputs for machine learning models. Nested cross validation was used for model selection, and a support vector classifier was identified as the most reliable framework for separating ductile and brittle alloys at room temperature. Model interpretation with Shapley values showed that the exchange–correlation parameter $r_s$, the valence electron concentration, the pressure field, and the electronegativity mismatch are the primary features controlling the decision boundary. \textcolor{red}{The learned empirical relationships between the physically-informed descriptors and ductile classifications were further examined using density functional theory calculations on representative alloys spanning the SVC decision boundary. The DFT-derived elastic descriptors, including the Pugh ratio and Poisson's ratio, showed broad agreement with the SVC decision-function margins. In particular, alloys assigned large positive margins generally exhibited mechanical characteristics associated with enhanced ductility, whereas compositions located near the decision boundary displayed more ambiguous behavior. Although the correspondence was not perfect, the DFT results provide independent physical support for the trends learned by the machine-learning model and suggest that the classifier captures meaningful relationships between alloy composition, bonding, and mechanical response.} While certain qualitative elemental tendencies such as Ti-assisted ductility are broadly consistent with metallurgical intuition, the present work provides a visualization of the multivariate decision boundary governing ductile versus brittle behavior in tungsten-containing RHEAs. By interpolating within the feature space and projecting the SVC decision function onto pseudo ternary diagrams, we visualized how the predicted likelihood of ductility evolves with composition and identified clear alloying strategies.
\textcolor{red}{This work establishes a physics-informed and interpretable framework for screening ductility in tungsten-containing refractory high-entropy alloys using sparse experimental data. The combination of composition-based descriptors, nested cross-validation, decision-boundary analysis, and first-principles validation provides a systematic approach for identifying alloying strategies associated with improved room-temperature ductility. While the present study focuses on room-temperature mechanical behavior as an initial screening criterion, the methodology is readily extensible to additional performance objectives relevant to fusion materials. Future work should integrate targeted experimental validation of the predicted alloy chemistries, expanded datasets generated under consistent testing protocols, and higher-fidelity electronic descriptors capable of more directly capturing the mechanisms governing bonding, dislocation mobility, and fracture. More broadly, the integration of interpretable machine learning with density functional theory provides a practical pathway for accelerating the discovery of tungsten-based alloys with reduced ductile-to-brittle transition temperatures. Comprehensive qualification for fusion applications will ultimately require simultaneous consideration of irradiation-induced property evolution, thermal transport, activation behavior, and thermodynamic stability within a multi-objective alloy design framework.}

\section*{Code Availability}

All machine learning models, alloy-design-space generators, prediction scripts, and figure-generation workflows developed in this study are openly available at:

\url{https://github.com/mahata-lab/Tungsten-Ductile-Brittle-Temperature}

The density functional theory workflow, including SQS generation, VASP input files, output files, and post-processing results, is available at:

\url{https://zenodo.org/records/20835365}

Together, these repositories provide the datasets, source code, simulation inputs, and analysis scripts required to reproduce all figures, tables, machine-learning results, and DFT calculations reported in this study.

%%%%%%%%%%%%%%%%%%%%%%%%%%%%%%%%%%%%%%%%%%%%%%%%%%

\section*{Acknowledgments}

This work was supported by the Department of Mechanical and Electrical Engineering at Merrimack College. This research also benefited from high performance computing allocations provided by the National Science Foundation through ACCESS (award MAT250103). Additional computational resources were supported by Argonne National Laboratory under the Director's Discretionary allocation for the project \textit{GNNMD}. Further support was provided through a National Science Foundation MRI Award to Wilkes University (Award No.\ 1920129), which contributed essential computational infrastructure for this study.

\bibliography{references}

\end{document}